\documentclass[letterpaper,dvipsnames]{vldb}
\usepackage[T1]{fontenc}
\usepackage[utf8]{inputenc}
\usepackage{array}
\usepackage{refstyle}
\usepackage{float}
\usepackage{booktabs}
\usepackage{calc}
\usepackage{url}
\usepackage{multirow}
\usepackage{amsmath}
\usepackage{amssymb}
\usepackage{graphicx}
\usepackage{wasysym}
\usepackage[unicode=true,
 bookmarks=true,bookmarksnumbered=true,bookmarksopen=false,
 breaklinks=false,pdfborder={0 0 1},backref=false,colorlinks=false]
 {hyperref}

\makeatletter

\AtBeginDocument{\providecommand\tabref[1]{\ref{tab:#1}}}
\AtBeginDocument{\providecommand\subsecref[1]{\ref{subsec:#1}}}
\AtBeginDocument{\providecommand\secref[1]{\ref{sec:#1}}}
\AtBeginDocument{\providecommand\figref[1]{\ref{fig:#1}}}
\AtBeginDocument{\providecommand\algref[1]{\ref{alg:#1}}}
\pdfpageheight\paperheight
\pdfpagewidth\paperwidth

\providecommand{\tabularnewline}{\\}
\floatstyle{ruled}
\newfloat{algorithm}{tbp}{loa}
\providecommand{\algorithmname}{Algorithm}
\floatname{algorithm}{\protect\algorithmname}
\RS@ifundefined{subsecref}
  {\newref{subsec}{name = \RSsectxt}}
  {}
\RS@ifundefined{thmref}
  {\def\RSthmtxt{theorem~}\newref{thm}{name = \RSthmtxt}}
  {}
\RS@ifundefined{lemref}
  {\def\RSlemtxt{lemma~}\newref{lem}{name = \RSlemtxt}}
  {}

\usepackage{graphicx}
\usepackage{balance}
\usepackage[noend]{algpseudocode}
\usepackage{semantic}
\usepackage{cite}
\usepackage{tikz}
\usepackage{tikzit}
\usepackage{booktabs}
\usepackage{xcolor}
\usepackage{soul}
\usepackage{pgfplots}
\pgfplotsset{compat=newest}
\pgfplotsset{table/search path={images}}
\usepackage[labelfont=bf]{caption}
\usetikzlibrary{arrows.meta}
\usetikzlibrary{calc}

\tikzstyle{arrow}=[-{Latex[length=3mm]}]

\renewcommand{\figref}[1]{Figure \ref{fig:#1}}
\renewcommand{\algref}[1]{Algorithm \ref{alg:#1}}
\renewcommand{\tabref}[1]{Table \ref{tab:#1}}
\renewcommand{\secref}[1]{\S~\ref{sec:#1}}
\renewcommand{\subsecref}[1]{\S~\ref{subsec:#1}}

\newcommand{\added}[1]{#1}
\newcommand{\deleted}[1]{{\color{OrangeRed}{\st{#1}}}}
\renewcommand{\deleted}[1]{}
\newcommand{\modified}[2]{\deleted{#1}\added{#2}}

\hypersetup{
    colorlinks,
    linkcolor={black},
    citecolor={black},
    urlcolor={black},
}
\makeatother

\begin{document}
\vldbTitle{Efficient Oblivious Database Joins}
\vldbAuthors{Simeon Krastnikov, Florian Kerschbaum, Douglas Stebila}
\vldbDOI{\url{https://doi.org/10.14778/3407790.3407814}}
\vldbVolume{13}
\vldbNumber{11}
\vldbYear{2020}
\vldbPages{2132-2145}
\numberofauthors{3}
\pagestyle{empty}
\title{Efficient Oblivious Database Joins}
\author{\alignauthor
Simeon Krastnikov\\
	\affaddr{University of Waterloo}\\
	\email{skrastnikov@uwaterloo.ca}
\alignauthor
Florian Kerschbaum\\
	\affaddr{University of Waterloo}\\
	\email{fkerschb@uwaterloo.ca}
\alignauthor
Douglas Stebila\\
	\affaddr{University of Waterloo}\\
	\email{dstebila@uwaterloo.ca}}
\maketitle
\begin{abstract}
A major algorithmic challenge in designing applications intended for
secure remote execution is ensuring that they are  oblivious to their
inputs, in the sense that their memory access patterns do not leak
sensitive information to the server. This problem is particularly
relevant to cloud databases that wish to allow queries over the client's
encrypted data. One of the major obstacles to such a goal is the join
operator, which is non-trivial to implement obliviously without resorting
to generic but inefficient solutions like Oblivious RAM (ORAM). 

We present an oblivious algorithm for equi-joins which (up to a logarithmic
factor) matches the optimal $O(n\log n)$ complexity of the standard
non-secure sort-merge join (on inputs producing $O(n)$ outputs).
We do not use use expensive primitives like ORAM or rely on unrealistic
hardware or security assumptions. Our approach, which is based on
sorting networks and novel provably-oblivious constructions, is conceptually
simple, easily verifiable, and very efficient in practice. Its data-independent
algorithmic structure makes it secure in various different settings
for remote computation, even in those that are known to be vulnerable
to certain side-channel attacks (such as Intel SGX) or with strict
requirements for low circuit complexity (like secure multiparty computation).
We confirm that our approach is easily realizable by means of a compact
\deleted{prototype }implementation which matches our expectations
for performance and is shown, both formally and empirically, to possess
the desired security characteristics.
\end{abstract}

\section{Introduction}

\begin{table*}
\begin{centering}
\caption{\textbf{Comparison of approaches for oblivious database joins.} $n_{1}$
and $n_{2}$ are the input table sizes, $n=n_{1}+n_{2}$, $m$ is
the output size, $m'=m+n_{1}+n_{2}$, $t$ is the amount of\emph{
}memory assumed to be oblivious. The time complexities are in terms
of the number of database entries and assume use of a bitonic sorter
for oblivious sorting (where applicable). \label{tab:related-work}}
\par\end{centering}
\centering{}%
\begin{tabular}{cccc}
\toprule 
\textbf{Algorithm/System} & \textbf{Time complexity} & \textbf{Local Memory} & \textbf{Assumptions/Limitations}\tabularnewline
\midrule 
Standard sort-merge join & $O(m'\log m')$ & $O(1)$ & not oblivious\tabularnewline
Agrawal et al. \cite{agrawal2006sovereign} (Alg. 3) & $O(n_{1}n_{2})$ & $O(1)$ & insecure (see \S~ 2.3.1 of \cite{li2008privacy})\tabularnewline
Li and Chen \cite{li2008privacy} (Alg. A2) & $O(mn_{1}n_{2}/t)$ & $O(t)$ & --\tabularnewline
Opaque \cite{zheng2017opaque} and ObliDB \cite{eskandarian2019oblidb} & $O(n\log^{2}(n/t))$ & $O(t)$ & restricted to primary-foreign key joins\tabularnewline
Oblivious Query Processing \cite{arasu2013oblivious} & $O(m'\log^{2}m')$ & $O(\log m')$ & missing details; performance concerns\tabularnewline
Ours & $O(m'\log^{2}m')$ & $O(1)$ & --\tabularnewline
\bottomrule
\end{tabular}
\end{table*}

With an increasing reliance on cloud-based services to store large
amounts of user data securely, there is also a growing demand for
such services to provide remote computation in a privacy-preserving
manner. This is a vital requirement for cloud databases that store
sensitive records and yet wish to support queries on such data. 

Various different mechanisms exist that achieve this purpose, for
instance, dedicated hardware in the form of secure cryptographic coprocessors
or hardware enclaves like Intel SGX \cite{costan2016intel} that come
in the form of a dedicated set of processor instructions. Although
such approaches provide good cryptographic guarantees in that they
ensure the contents of the user's data remain encrypted throughout
the execution of a remote program, on their own they provide no guarantees
to address a major source of information leakage: the memory access
patterns of the execution. As the program reads and writes to specific
addresses of the untrusted server's memory, these access patterns
can reveal information to the server about the user's data if the
program's control flow is dependent on its input. 

Consider, for example, the standard $O(n\log n)$ sort-merge algorithm
for database joins. The two input tables are first sorted by their
join \modified{attribute}{attribute values} and then scanned for
matching entries by keeping track of a pointer at each table. At each
step either one of the pointers is advanced (if one of either corresponding
join \modified{attributes}{attribute values} precedes the other),
or an entry is appended to the output. If the tables are stored in
regular memory, an adversary observing memory patterns will obtain
input-dependent information at each step. Namely, at each step it
will learn the locations of the two entries read from the input table,
and depending on whether an entry is written to output, it will learn
whether the entries match. This information can reveal critical information
about the user's input.

Protecting a program against such leaks amounts to making it \emph{oblivious}:
that is, ensuring that its decisions about which memory locations
to access do not depend on the contents of its input data. For many
programs this is difficult to accomplish without introducing substantial
computational overhead \cite{zahur2013circuit}. The generic approach
is to use an Oblivious RAM (ORAM), which provides an interface through
which a program can access memory non-obliviously, while at the same
time providing guarantees that the physical accesses of such programs
are oblivious. Though the design of ORAM schemes has been a central
focus of oblivious algorithm design, such schemes have a very high
computational overhead, not only due to their asymptotic overhead
(due to a theoretical lower bound of $O(\log n)$ time per access
to an array of $n$ entries),  but also due to the fact that they
can be inefficient in practice \cite{zahur2013circuit,islam2012access,hoang2017s3oram,stefanov2014practical}
and in some cases -- insecure \cite{abraham2017asymptotically}.

Another approach to achieving obliviousness is to assume a limited
but non-constant amount of memory that can be accessed non-obliviously.
While such an assumption may make sense in certain settings (for example,
cryptographic coprocessors may provide internal memory protected from
the untrusted system), it is unsafe to make in the more common hardware
enclave setting due to a wide range of attacks that can infer access
patterns to enclave memory itself \cite{brasser2017software,gotzfried2017cache,lee2017inferring,wang2017leaky,xu2015controlled,van2017telling}.

These considerations motivate the need for the design of problem-specific
oblivious algorithms that closely approach the efficiency of their
non-secure counterparts without the use of generic primitives or reliance
on hardware assumptions. Such algorithms are very similar to circuits
in that their control flow is independent of their inputs. This not
only provides security against many side-channel attacks (beyond those
involving memory accesses), but also makes them very suitable for
use in secure multiparty computation where programs with low circuit
complexity achieve the best performance \cite{zahur2016revisiting,wang2015circuit}.
One of the best-known oblivious algorithms are sorting networks such
as those proposed by Batcher \cite{batcher1968sorting}, which match
(up to at most a logarithmic factor) the standard $O(n\log n)$ complexity
of sorting and are often a critical component in other oblivious algorithms
such as ours. 

Making database operators oblivious does not pose much of an algorithmic
challenge in most cases since often one can directly apply sorting
networks (for instance to select or insert entries). On the other
hand, database joins, being among the most algorithmically-complex
operators, have proven to be very difficult to make oblivious in the
general case. This is due to the fact that one cannot allow any such
oblivious algorithm to base its memory accesses on the structure of
the input tables (with respect to how many entries from one table
match a given entry in the other table). As such, joins have been
the prime focus of work on oblivious database operators: prior work
is summarized in \tabref{related-work} and discussed in detail in
\subsecref{related-work}.

\subsubsection*{Contributions}

We fully describe an oblivious algorithm for \added{binary} database
equi-joins that achieves $O(n\log^{2}n+m\log m)$ running time, where
$n$ is the input length (the size of both tables) and $m$ is the
output length, thus matching the running time of the standard non-oblivious
sort-merge join up to a logarithmic factor. It can also achieve a
running time of $O(n\log n+m\log m)$ but only using a sorting network
that is too slow in practice. Our algorithm does not use ORAM or any
other computationally-expensive primitives and it does not make any
hardware assumptions other than the requirement of a constant-size
working set of memory (on the order of the size of one database entry),
for example to compare two entries. In other words, any model of computation
that can support a sorting network on encrypted data can also support
our algorithm; this includes secure coprocessors or hardware enclaves
like Intel SGX, in which case we provide resistance against various
side-channel attacks. In addition, because our program is analogous
to a circuit, it is very suitable for use in settings like secure
multiparty computation and fully homomorphic encryption.

\modified{Our approach is conceptually very simple (we have a working
prototype consisting of just 600 lines of C++ code), being based on
a few repeated runs of a sorting network and other basic primitives.
As such, it is both efficient, amenable to a certain degree of parallelization,
and easy to verify for obliviousness. We have used a dedicated type
system to verify obliviousness of the prototype implementation itself
and have conducted experiments that empirically examine its memory
accesses and runtime.}{Our approach is conceptually very simple,
being based on a few repeated runs of a sorting network and other
basic primitives. As such it is both incredibly efficient, amenable
to a high degree of parallelization, and fairly easy to verify for
obliviousness. We have a working prototype implementation consisting
of just 600 lines of C++ code, as well as a version that makes use
of SGX. We have used a dedicated type system to formally verify the
obliviousness of the implementation and have conducted experiments
that empirically examine its memory accesses and runtime.}

\subsubsection*{Outline}

We begin by providing, in \secref{modes-computation}, a brief background
on several different modes of computation that our algorithm is compatible
with. In \secref{oblivious-programs}, we discuss the goals, challenges,
and methodologies related to oblivious algorithm design. \secref{problem-overview}
describes the target problem and prior related work, as well as give
the intuition behind our approach. The full algorithm is then described
in detail in \secref{algorithm-description}. Finally, we discuss
our implementation in \secref{evaluation}, where we also analyze
its security and performance.

\section{Computing on Encrypted Data}

\label{sec:modes-computation}Users who have securely stored their
data on a remote server often need to perform computation on their
encrypted data, for example, to execute database queries. We discuss
and contrast several different approaches that strive to achieve this
purpose.

\subsubsection*{Outsourced External Memory}

In this setting (discussed in \cite{goodrich2014data}), there is
no support for server-side computation on the client's private data.
The client treats the server as external memory: if they wish to compute
on their data, they must do so locally. This scenario is clearly impractical
for intensive computations due to the significant differences between
RAM and network latency. 

\subsubsection*{Secure Cryptographic Coprocessors}

A cryptographic coprocessor (e.g., \cite{arnold2012ibm}) is a tamper-proof
device can perform computation within its own trusted region isolated
from its external host. Through the use of remote attestation, the
client can send a trusted code base (TCB) to the coprocessor and have
it execute within a secure environment shielded from the semi-trusted
server (according to the Trusted Platform Module specification \cite{iso_central_secretary_systems_2016}).
The drawback to this approach is that such hardware provides very
limited memory and computational power, and imposes hardware requirements
on the server. 

\subsubsection*{Trusted Execution Environments (TEE)}

\emph{Hardware enclaves} such as Intel SGX \cite{costan2016intel},
provide similar guarantees to coprocessors in that the client can
make use of remote attestation to run a TCB within a trusted execution
environment (TEE) that provides guarantees for authenticity and some
protection against an untrusted OS. Such designs are becoming increasingly
prevalent in new processors, taking the form of a specialized set
of instructions for setup and access to the TEE. The enclave provides
a limited amount of memory called the Enclave Page Cache (EPC), which
resides on the system's main memory but cannot be accessed by other
processes (including the kernel). In addition, the contents of the
enclave are encrypted, and the processor can seamlessly read, write
and perform logical and arithmetic operations on this data. Although
these properties make it seem that hardware enclaves provide a secure
container completely isolated from the untrusted OS, numerous papers
\cite{brasser2017software,gotzfried2017cache,lee2017inferring,wang2017leaky,xu2015controlled,van2017telling}
have shown that enclaves like Intel SGX are susceptible to numerous
side-channel attacks, e.g., cache attacks that infer data-dependent
information based on memory access patterns to enclave memory.

\subsubsection*{Secure Multiparty Computation (SMC)}

In the general setting, secure multiparty computation \cite{goldreich2009foundations}
allows several parties to jointly compute \emph{functionalities }on
their secret inputs without revealing anything more about their inputs
than what can be inferred from the output. The two standard approaches
are Yao's garbled circuit protocol \cite{yao1986generate}, and the
Goldreich-Micali-Wigderson protocol \cite{goldreich2019play}, based
on secret sharing. Both approaches require the desired functionality
to be expressed as a boolean circuit, and the output is computed gate
by gate. Practical implementations of SMC include the SCALE-MAMBA
system \cite{aly2019scale}, as well as the ObliVM framework \cite{liu2015oblivm},
which allows programs to be written in a (restricted) high-level syntax
that can then be compiled to a circuit. Outsourcing computation using
SMC is usually done using a distributed protocol involving a cluster
of several servers \cite{bater2017smcql}.

\subsubsection*{Fully Homomorphic Encryption (FHE)}

Cryptosystems such as that of Gentry et al. \cite{gentry2009fully}
allow arbitrary computation on encrypted data. As in SMC, such schemes
require the target computation to be represented as a boolean circuit.
Although FHE provides solid theoretical guarantees and several implementations
already exist, it is currently too computationally-expensive for practical
use. 

\section{Oblivious Programs}

\begin{table}
\caption{\textbf{Properties of three levels of obliviousness.} Bottom portion
of table shows vulnerability of programs satisfying these levels to
timing (\emph{t}), page access attacks on data (\emph{pd}), page access
attacks on code (\emph{pc}), cache-timing (\emph{c}), or branching
(\emph{b}) attacks when used in different settings.\label{tab:security-levels}}

\centering{}%
\begin{tabular}{cccc}
\toprule 
\textbf{Property/Setting} & \multicolumn{1}{c}{\textbf{I}} & \multicolumn{1}{c}{\textbf{II}} & \multicolumn{1}{c}{\textbf{III}}\tabularnewline
\midrule
\modified{Local memory}{Constant local memory} & $\times$ & $\checked$ & $\checked$\tabularnewline
Circuit-like & $\times$ & $\times$ & $\checked$\tabularnewline
\midrule
Ext. Memory & \emph{t} & \emph{t} & $\checked$\tabularnewline
Secure Coprocessor & \emph{t} & \emph{t} & $\checked$\tabularnewline
TEE (enclave) & \emph{\hspace{-0.6em}t, pd, pc, c, b} & \emph{t, pc, c, b} & $\checked$\tabularnewline
Secure Computation & n/a & n/a & $\checked$\tabularnewline
FHE & n/a & n/a & $\checked$\tabularnewline
\bottomrule
\end{tabular}
\end{table}

\label{sec:oblivious-programs}Intuitively, a program is \emph{data-oblivious
}(or simply \emph{oblivious)} if its control flow, in terms of the
memory accesses it makes, is independent of its input given the input
size. That is, for all inputs of the same length, the sequence of
memory accesses made by an oblivious program is always identical (or
identically distributed if the program is probabilistic). This is
indeed well-defined for simple computational models like random-access
machines and Turing machines; for instance, the latter is said to
be oblivious if the motions of its head are independent of its input.
However, we need to carefully account for real-world hardware where
there can be different types of memory as well as various side-channels
unaccounted for in simpler models.

In the remainder of this section, we define our adversarial model,
and introduce three different levels of obliviousness that one can
obtain against such an adversary. We also introduce various tools
and methodologies related to obliviousness.

\subsection{Adversarial Model}

We will use an abstract random access machine model of computation
where we distinguish between two types of memory. During an execution
of a program, the adversary has complete view and control of the \emph{public}
\emph{memory }(for example, RAM) used throughout its execution. However,
the program may use a small amount of \emph{local memory }(or \emph{protected
memory}) that is completely hidden from the adversary (for example,
processor registers). The program may use this memory to perform computations
on small chunks of data; the adversary learns nothing about such computations
except the time spent performing them (we assume that all processor
instructions involving local memory that are of the same type take
an equal amount of time).

We assume that the adversary cannot infer anything about the individual
contents of individual cells of public memory, as well as whether
the contents of a cell match a previous value. This can be achieved
through the use of a probabilistic encryption scheme and is not the
concern of this paper.

\subsection{Degrees of Obliviousness\label{subsec:types-obliviousness}}

We distinguish between three different levels of obliviousness that
a given program may satisfy (summarized in \tabref{security-levels}),
from weakest to strongest, each subsuming the lower levels. The distinctions
will be based on how much local memory the program assumes and whether
the program's use of local memory leaks information through side-channels.
We will restrict our attention to deterministic programs; the concepts
easily generalize to the probabilistic case.

\subsubsection*{Level I}

A program is oblivious in this sense if its accesses to public memory
are oblivious but it requires a non-constant amount \deleted{($\omega(1)$
with regards to the input size)} of local memory used for non-oblivious
computation\deleted{ (though the amount of local memory is still
assumed to be relatively small compared to to the amount of public
memory since otherwise we would always use only local memory)}. This
memory may be accessed whenever required and for any duration of time.
Such programs are suitable for use in the outsourced external memory
model since the client can use as much local memory as there is memory
on his machine. They may also be suitable for use in a secure coprocessor
model setting since coprocessors have an internal memory separated
from the rest of the system. However, in both of these scenarios,
timing attacks may be an issue: e.g., if the local memory is used
for variable lengths of time between pairs of public accesses. 

Examples of algorithms that are oblivious in this sense are those
proposed by Goodrich \cite{goodrich2014data,eppstein2010privacy,goodrich2011data},
which are well-suited for the outsourced external memory model.

\subsubsection*{Level II}

At this level we not only require, as before, accesses to public memory
to be oblivious but also that \modified{the program use a constant
amount of local memory}{the amount of local memory used by the program
is bounded by a constant}. \modified{In practice, this means that
the program can only access RAM obliviously but may use the registers
and cache of the processor to perform a series of plaintext computations,
for example, to compute the condition for a branch or to perform an
arithmetic operation on two words read (obliviously) from memory.
}{In practice, the exact size of this constant  depends on the amount
of available CPU register and cache memory, which can be used for
example, to compute the condition for a branch or to perform an arithmetic
operation on two words read (obliviously) from RAM. Any such accesses
must be on inputs that fit in one cache line so as to not cause non-oblivious
RAM accesses due to cache evictions. } \deleted{If the program needs
to perform operations on data that is too large to fit within the
processor's registers and cache, the data is split up into word-sized
chunks and a sequence of operations is performed on all of the chunks
(to maintain obliviousness).}

Making the distinction between this level and the previous is motivated
by the fact that hardware enclaves like Intel SGX are vulnerable to
side-channel attacks based on page-level accesses patterns to enclave
memory itself \cite{xu2015controlled,van2017telling}, which have
been shown to be extremely powerful, often succeeding in extracting
sensitive data and even whole files. Therefore one cannot assume that
the Enclave Page Cache provides oblivious memory. In works like Oblix
\cite{mishra2018oblix} level II programs are called \emph{doubly-oblivious
}since, in the hardware enclave context, their accesses to both regular
and enclave memory are oblivious.

Although it may seem that a level II program is safe against the above
attacks, this is not quite the case, though it certainly fares better
in this respect than a level I program. The \emph{data} of the program
will be accessed obliviously, but its actual machine code, which is
stored in memory, will be accessed based on the control flow of the
program. The program may branch in a data-dependent way and though
the memory accesses to public data in both branches are required to
be the same, each branch will access a different fragment of the program's
machine code, thus leaking information about the data that was branched
on.

\subsubsection*{Level III}

This is a strong notion of obliviousness where we require that the
control flow of the program, down to the level of the exact processor
instructions it executes, be completely independent of its input,
except possibly its length. In other words, the program counter has
to always go through the same sequence of values for all inputs of
the same length. We can think of such a program as a family of circuits,
one for each input size; as such, it is very well-suited for secure
multiparty computation (and fully homomorphic encryption).

This definition is also motivated by the fact that additional measures
are required to provide protection against attacks based on accesses
to machine code as well as the fact that hardware enclaves have also
been shown to be vulnerable to a variety of other side-channel attacks
such as cache-timing \cite{brasser2017software,wang2017leaky,gotzfried2017cache},
branching \cite{lee2017inferring}, or other types of timing attacks.
Such attacks can infer data based on the control flow of a program
at the instruction level: this includes the way it accesses the registers
and cache of the processor, as well as the exact number of instructions
it performs. A level III program will be secure against these attacks
so long as care is taken to prevent the compiler from introducing
data-dependent optimizations. \added{(Many compilers such as GCC
support selectively chosen optimizations, but the specifics of using
this approach for oblivious programs is  outside the scope of this
paper.)}

\subsubsection*{Revealing Output Length}

By producing an output of length $m$, a program reveals data-dependent
information about the input. We can always eliminate this problem
by padding the output to its maximum possible size; however, this
can result in suboptimal running time. For instance, the join operator
can produce an output of up to $O(n^{2}$) on an input of size $n$,
which means that any join algorithm that pads its output must have
at least quadratic runtime. For this reason, we will only consider
programs that do not pad their output and thus leak the output size
$m$\added{, as well as their runtime}.

\subsection{Oblivious RAM (ORAM)}

The most general approach to making arbitrary programs oblivious (in
any of the above senses) is to use an Oblivious RAM (ORAM), a primitive
first introduced by Goldreich and Ostrovsky \cite{goldreich1987towards,goldreich1996software}.
An ORAM simulates a regular RAM in such a way that its apparent physical
memory accesses are independent of those being simulated, thus providing
a general approach to compiling general programs to oblivious ones.
In other words, by using an ORAM as an interface through which we
store and access sensitive data, we can eliminate access pattern leaks,
though in doing so we incur at least a logarithmic overhead per memory
access according to the Goldreich-Ostrovsky lower bound \cite{goldreich1996software}.
Even if such overhead is acceptable in terms of the overall asymptotic
complexity, ORAM constructions tend to have prohibitively large constant
overhead, which make them impractical to use on reasonably-sized inputs.

One of the well-known ORAM schemes is Path ORAM \cite{stefanov2013path},
which produces programs that satisfy level I obliviousness (Oblix
\cite{mishra2018oblix} gives a modification that is oblivious at
level II). Various schemes have been introduced that make ORAM more
suitable for SMC by optimizing its resulting circuit complexity (i.e.,
how close it is to producing a level III program) \cite{wang2015circuit,zahur2016revisiting,gentry2013optimizing,doerner2017scaling}.
Despite the abundance of ORAM schemes, their high performance cost
\cite{zahur2013circuit,islam2012access,hoang2017s3oram,stefanov2014practical}
(due to their polylogarithmic complexity overhead, their large hidden
constants, and issues with parallelizability), calls for a need for
problem-specific oblivious algorithm design.

\subsection{From Oblivious Programs to Circuits\label{subsec:obliv-to-circuit}}

Given a program that satisfies level II obliviousness, approaches
exist to transform it into a a circuit-like level III program while
introducing only a constant overhead \cite{rane2015raccoon,molnar2005program,liu2015oblivm,liu2013memory}.
There are three  additional constraints that the control flow of the
program must satisfy for this to be the case:

\emph{1.} Any loop condition must depend on either a constant or the
input size. This corresponds to the fact that all loops must be unrolled
if one wishes to obtain a literal boolean circuit. For instance if
$secret$ is a variable that depends on the contents of the input
data, we cannot allow behaviour like:

\medskip{}

\noindent\fbox{\begin{minipage}[t]{1\columnwidth - 2\fboxsep - 2\fboxrule}%
\begin{algorithmic}

\State $i \gets 0$
\While{$i < secret$}
	\State $i \gets i + 1$
\EndWhile

\end{algorithmic}%
\end{minipage}}

\medskip{}

Though this code will make no memory accesses if the $i$ counter
is stored in a register, it is very hard to automatically protect
such code against timing attacks in general (though in this case the
fix is obvious: replace the while loop with $i\gets secret$). 

\emph{2.} The branching depth of the program --- the maximum number
of conditional branches encountered by any given run --- is constant.
This requirement allows us to eliminate conditional statements without
affecting runtime complexity.  A statement like

\medskip{}

\noindent\fbox{\begin{minipage}[t]{1\columnwidth - 2\fboxsep - 2\fboxrule}%
\begin{algorithmic}

\If {$secret$}
	\State $x_1 \gets y_1$
	\State $x_3 \gets y_3$
\Else
	\State $x_1 \gets z_1$
	\State $x_2 \gets z_2$
\EndIf

\end{algorithmic}%
\end{minipage}}\medskip{}
can be replaced by

\medskip{}

\noindent\fbox{\begin{minipage}[t]{1\columnwidth - 2\fboxsep - 2\fboxrule}%
\begin{algorithmic}

\State $x_1 \gets y_1 \cdot secret + z_1 \cdot (\neg secret)$
\State $x_2 \gets z_2 \cdot 0 + z_2 \cdot (\neg secret)$
\State $x_3 \gets y_3 \cdot secret + y_3 \cdot 0$

\end{algorithmic}%
\end{minipage}}

\medskip{}

This increases the total computation by a factor of 2. On the other
hand, if we have a sequence of $d$ nested conditional statements,
the computational overhead will be on the order of $2^{d}$, which
is why we require $d$ to be constant.

\emph{3. }If the program reveals the output length $m$, it does so
only after allocating $m_{0}\in\Omega(m)$ memory. This is so the
level II program can be split into two circuit-like level III programs
that are to be run in sequence: one parameterized by $n$ that computes
the value of $m_{0}$, and a second parameterized by both $n$ and
$m_{0}$. 

\subsection{Oblivious Sorting}

\begin{sloppypar}Sorting networks such as bitonic sorters \cite{batcher1968sorting}
provide an in-place input-independent way to sort $n$ elements in
$O(n\log^{2}n)$ time, taking the form of an $O(\log^{2}n)$-depth
circuit. Although $O(n\log n)$ constructions also exist, they are
either very inefficient in practice (due to large constant overheads)
or non-parallelizable \cite{goodrich2014zig}. Each non-recursive
step of a bitonic sorter reads two elements at fixed input-independent
locations, runs a comparison procedure between the two elements and
swaps the entries depending on the result. To ensure obliviousness,
even if the elements are not to be swapped, the same (re-encrypted)
entries are written to their original locations. When a probabilistic
encryption scheme is used, this leaks no information about whether
the two elements were swapped. \end{sloppypar}

We parameterize our calls to a bitonic sorter with a lexicographic
ordering on chosen element attributes. For example, if $A$ is a list
of elements where each element has attributes $x,y,z,\ldots$, then
\[
\text{\textsc{Bitonic-Sort}\ensuremath{\langle x\uparrow,y\uparrow,z\downarrow\rangle(A)}}
\]
will sort the elements in \texttt{$A$} by increasing $x$ attribute,
followed by increasing $y$ attribute, and then by decreasing $z$
attribute.

We can use sorting networks as filters. For instance, we will use
$\varnothing$ to designate a void (null) entry that is marked to
be discarded (or often a ``dummy'' entry) so that if we know that
\texttt{$A$} of size $n$ has $k$ non-null elements, we can run
\[
\text{\textsc{Bitonic-Sort}\ensuremath{\langle\neq\varnothing\uparrow\rangle}}(A)
\]
and collect the first $k$ non-null elements in the output. Alternatively,
Goodrich \cite{goodrich2011data} has proposed an efficient $O(n\log n)$
oblivious algorithm specifically for this problem (there referred
to as \emph{compaction}).

\section{Problem Overview }

\label{sec:problem-overview}In this section, we describe the general
problem, the security goals our solution will satisfy, and prior work
in similar directions. We then briefly outline the general idea behind
our approach.

\subsection{Problem Definition}

We are given as input two unsorted tables and are interested in computing
the \added{binary} equi-join of both tables (though the ideas extend
to \added{some} more general types of inner joins). That is, our
input consists of two tables $T_{1}$ and $T_{2}$, each consisting
of respectively $n_{1}$ and $n_{2}$ (possibly-repeated) pairs $(j,d)$\emph{
}\modified{(we call $j$ a \emph{join attribute }and $d$ a \emph{data
attribute})}{(we call $j$ a \emph{join (attribute) value} and $d$
a \emph{data (attribute) value})}\emph{. }The output we would like
to compute is
\[
T_{1}\bowtie T_{2}=\{(d_{1},d_{2})\mid(j,d_{1})\in T_{1},\ (j,d_{2})\in T_{2}\}.
\]
 The tables $T_{1}$, $T_{2}$, and $T_{1}\bowtie T_{2}$ are not
assumed to be ordered\deleted{, though in our case the output will
be sorted lexicographically, by $(j,d)$}.

\subsection{Related Work\label{subsec:related-work}}

The design of oblivious join operators has been studied both in isolation
and also as part of larger privacy-oriented database systems, where
it is emphasized as the most challenging component; we compare several
different approaches in \tabref{related-work}. Agrawal et al. \cite{agrawal2006sovereign}
propose several join algorithms for use in a setting similar to ours;
however, their roughly $O(n_{1}n_{2})$ complexity is close to that
of a trivial $O(n_{1}n_{2}\log^{2}(n_{1}n_{1}))$ oblivious algorithm
based on a nested loop join. Additionally, their security definition
allows leakage of a certain property of the inputs (as pointed out
in \cite{li2008privacy}, where the issue was fixed without significant
improvements in runtime). SMCQL \cite{bater2017smcql} is capable
of processing SQL queries through secure computation primitives but
its secure join also runs in $O(n_{1}n_{2})$ time. \added{Conclave
\cite{volgushev2019conclave} implements join operators for SMC; however,
its approach involves revealing entries to a ``selectively-trusted
party''.}

Opaque, which is geared towards private database queries in a distributed
setting, implements an oblivious sort-merge algorithm \cite{zheng2017opaque}
(as well as its variant in ObliDB \cite{eskandarian2019oblidb}) but
handles only the specific case of primary-foreign key joins (in which
case $m=O(n)$ and their $O(n\log^{2}(n/t))$ complexity matches ours
for constant $t$). Though Opaque makes use of the $O(t)$ available
enclave memory to optimize its running time, such optimizations rely
on ``enclave designs that protect against access patterns to the
EPC'' such as ``Sanctum, GhostRider, and T-SGX'' to obtain a pool
of oblivious memory (meaning that the optimized versions are only
level I oblivious). Such constructions could potentially introduce
a computational overhead that outweigh any optimizations (GhostRider
for instance relies on ORAM) or introduce additional hardware requirements
and security assumptions. 

The closest to our work is that of Arasu et al. \cite{arasu2013oblivious},
which mirrors the overall structure of our algorithm but ultimately
reduces to a different (arguably more challenging) problem than the
one we deal with (``obliviously reordering {[}sequences{]} to make
{[}them{]} barely prefix heavy''). The details for the proposed solution
to this problem are incomplete and the authors have not provided a
proof-of-concept implementation that shows empirical results. We believe
that even if a solution to this problem exists, the overall algorithm
will be less efficient than ours due to the high constant overheads
from repeated sorts. Lastly, their approach also assumes\added{, by default,}
a local memory of $O(\log(m+n))$ entries (thus being level I oblivious)
since it is intended for use in a secure coprocessor setting. For
use in more practical settings like Intel SGX, such memory would either
have to be obtained in the same manner as was argued above for the
case of Opaque or through an $O(\log(m+n))$ time complexity overhead
for each local memory access (that is, if each access achieves obliviousness
by reading all entries in local memory). 

\begin{sloppypar}Encrypted databases like CryptDB \cite{popa2011cryptdb}
employ deterministic and partial homomorphic encryption to process
databases in a hardware-independent way but such databases are non-oblivious.
\added{Private Set Intersection (PSI) and Private Record Linkage
(PRL) are somewhat similar problems to the one considered in this
paper in that they involve finding matching entries among different
databases. Although some protocols for these problems rely solely
on SMC techniques (by constructing circuits as in our work), more
efficient protocols make use of cryptographic primitives like oblivious
transfer (extension) that are not applicable to database joins (see
the survey on PSI in \cite{pinkas2018scalable}). }\end{sloppypar}

\subsection{Security Characteristics}

Intuitively, our algorithm will be oblivious with regards to the way
it accesses any memory with non-constant size. More precisely, we
will provide security in the form of level II obliviousness, as described
in \subsecref{types-obliviousness}. Hence, our condition will be
that for all inputs of length $n$ that produce outputs of equal length
$m$, the sequence of memory accesses our algorithm makes on each
the inputs is always the same (or identically distributed if we make
use of randomness). 

We will use a constant amount of local memory on the order of the
size of a single database entry, which we will use to process entries
and keep counters. That is, our accesses to public memory (where the
the input, output and intermediate tables are stored) will be of the
form

\medskip{}

\noindent\fbox{\begin{minipage}[t]{1\columnwidth - 2\fboxsep - 2\fboxrule}%
\begin{algorithmic}

\State $e \overset{\star}{\gets} T[i]$
\State $\dots$
\State \emph{(sequence of operations on $e$)}
\State $\dots$
\State $T[i] \overset{\star}{\gets} e$

\end{algorithmic}%
\end{minipage}}\medskip{}

The notation $e\overset{\star}{\gets}T[i]$ explicitly signifies that
the $i$-th entry of the table $T$, which is stored in public memory,
is read into the variable $e$ stored in local memory. Our code will
be such that the memory trace consisting of all $\overset{\star}{\gets}$
operations (distinguished by whether they read or write to public
memory), are independent of the input sizes $n_{1}$ and $n_{2}$,
and the output size $m$.

As argued in \subsecref{obliv-to-circuit}, a level II program can
easily be transformed to a level III program with constant computational
overhead as long as no loop conditions depend on the input and the
branching depth is constant. Our approach will satisfy these properties
and thus yield a program that is secure against many of the side-channel
attacks listed in \subsecref{types-obliviousness}.

\subsection{Overview of Approach}

\begin{figure}
\resizebox{\columnwidth}{!}{
    \begin{tikzpicture}
        \tikzstyle{every node}=[font=\huge]
	    \begin{pgfonlayer}{nodelayer}
		    \node [style=none] (0) at (-5, 6) {};
		    \node [style=none] (1) at (-5, 0) {};
		    \node [style=none] (2) at (-2, 0) {};
		    \node [style=none] (3) at (-2, 6) {};
		    \node [style=none] (4) at (-4.25, 5.5) {$x$};
		    \node [style=none] (5) at (-4.25, 4.5) {$x$};
		    \node [style=none] (6) at (-4.25, 3.5) {$y$};
		    \node [style=none] (7) at (-4.25, 2.5) {$y$};
		    \node [style=none] (8) at (-4.25, 1.5) {$y$};
		    \node [style=none] (9) at (-2.75, 5.5) {$a_1$};
		    \node [style=none] (10) at (-2.75, 4.5) {$a_2$};
		    \node [style=none] (11) at (-2.75, 3.5) {$b_1$};
		    \node [style=none] (12) at (-2.75, 2.5) {$b_2$};
		    \node [style=none] (13) at (-2.75, 1.5) {$b_3$};
		    \node [style=none] (14) at (-3.5, -1) {$T_1$};
		    \node [style=none] (15) at (15, 6) {};
		    \node [style=none] (16) at (15, 0) {};
		    \node [style=none] (17) at (18, 0) {};
		    \node [style=none] (18) at (18, 6) {};
		    \node [style=none] (19) at (15.75, 5.5) {$x$};
		    \node [style=none] (20) at (15.75, 4.5) {$x$};
		    \node [style=none] (21) at (15.75, 3.5) {$x$};
		    \node [style=none] (22) at (15.75, 2.5) {$y$};
		    \node [style=none] (23) at (15.75, 1.5) {$y$};
		    \node [style=none] (24) at (17.25, 5.5) {$u_1$};
		    \node [style=none] (25) at (17.25, 4.5) {$u_2$};
		    \node [style=none] (26) at (17.25, 3.5) {$u_3$};
		    \node [style=none] (27) at (17.25, 2.5) {$v_1$};
		    \node [style=none] (28) at (17.25, 1.5) {$v_2$};
		    \node [style=none] (29) at (16.5, -1) {$T_2$};
		    \node [style=none] (30) at (-4.25, 6.75) {$j$};
		    \node [style=none] (31) at (-2.75, 6.75) {$d$};
		    \node [style=none] (37) at (0, 6) {};
		    \node [style=none] (38) at (0, -3) {};
		    \node [style=none] (39) at (3, -3) {};
		    \node [style=none] (40) at (3, 6) {};
		    \node [style=none] (41) at (0.75, 5.5) {$x$};
		    \node [style=none] (42) at (0.75, 4.5) {$x$};
		    \node [style=none] (43) at (0.75, 3.5) {$x$};
		    \node [style=none] (44) at (0.75, 2.5) {$x$};
		    \node [style=none] (45) at (0.75, 1.5) {$x$};
		    \node [style=none] (46) at (2.25, 5.5) {$a_1$};
		    \node [style=none] (47) at (2.25, 4.5) {$a_1$};
		    \node [style=none] (48) at (2.25, 3.5) {$a_1$};
		    \node [style=none] (49) at (2.25, 2.5) {$a_2$};
		    \node [style=none] (50) at (2.25, 1.5) {$a_2$};
		    \node [style=none] (51) at (1.5, -4) {$S_1$};
		    \node [style=none] (52) at (10, 6) {};
		    \node [style=none] (53) at (10, -3) {};
		    \node [style=none] (54) at (13, -3) {};
		    \node [style=none] (55) at (13, 6) {};
		    \node [style=none] (73) at (0.75, 0.5) {$x$};
		    \node [style=none] (74) at (0.75, -0.5) {$y$};
		    \node [style=none] (75) at (0.75, -1.5) {$y$};
		    \node [style=none] (76) at (2.25, 0.5) {$a_2$};
		    \node [style=none] (77) at (2.25, -0.5) {$b_1$};
		    \node [style=none] (78) at (2.25, -1.5) {$b_1$};
		    \node [style=none] (80) at (1.5, -2.5) {$...$};
		    \node [style=none] (99) at (-4.25, 0.5) {$y$};
		    \node [style=none] (100) at (-2.75, 0.5) {$b_4$};
		    \node [style=none] (101) at (15.75, 0.5) {$z$};
		    \node [style=none] (102) at (17.25, 0.5) {$w_1$};
		    \node [style=none] (103) at (15.75, 6.75) {$j$};
		    \node [style=none] (104) at (17.25, 6.75) {$d$};
		    \node [style=none] (105) at (0.75, 6.75) {$j$};
		    \node [style=none] (106) at (2.25, 6.75) {$d$};
		    \node [style=none] (107) at (10.75, 6.75) {$j$};
		    \node [style=none] (108) at (12.25, 6.75) {$d$};
		    \node [style=none] (109) at (5, 6) {};
		    \node [style=none] (110) at (5, -3) {};
		    \node [style=none] (111) at (8, -3) {};
		    \node [style=none] (112) at (8, 6) {};
		    \node [style=none] (113) at (6.5, -4) {$S_2$};
		    \node [style=none] (114) at (5.75, 5.5) {$x$};
		    \node [style=none] (115) at (5.75, 4.5) {$x$};
		    \node [style=none] (116) at (5.75, 3.5) {$x$};
		    \node [style=none] (117) at (5.75, 2.5) {$x$};
		    \node [style=none] (118) at (5.75, 1.5) {$x$};
		    \node [style=none] (119) at (7.25, 5.5) {$u_1$};
		    \node [style=none] (120) at (7.25, 4.5) {$u_2$};
		    \node [style=none] (121) at (7.25, 3.5) {$u_3$};
		    \node [style=none] (122) at (7.25, 2.5) {$u_1$};
		    \node [style=none] (123) at (7.25, 1.5) {$u_2$};
		    \node [style=none] (124) at (5.75, 0.5) {$x$};
		    \node [style=none] (125) at (5.75, -0.5) {$y$};
		    \node [style=none] (126) at (5.75, -1.5) {$y$};
		    \node [style=none] (127) at (7.25, 0.5) {$u_3$};
		    \node [style=none] (128) at (7.25, -0.5) {$v_1$};
		    \node [style=none] (129) at (7.25, -1.5) {$v_2$};
		    \node [style=none] (130) at (6.5, -2.5) {$...$};
		    \node [style=none] (131) at (5.75, 6.75) {$j$};
		    \node [style=none] (132) at (7.25, 6.75) {$d$};
		    \node [style=none] (133) at (8.5, 1.5) {};
		    \node [style=none] (134) at (9.5, 1.5) {};
		    \node [style=none] (135) at (10.75, 5.5) {$x$};
		    \node [style=none] (136) at (10.75, 4.5) {$x$};
		    \node [style=none] (137) at (10.75, 3.5) {$x$};
		    \node [style=none] (138) at (10.75, 2.5) {$x$};
		    \node [style=none] (139) at (10.75, 1.5) {$x$};
		    \node [style=none] (140) at (12.25, 5.5) {$u_1$};
		    \node [style=none] (141) at (12.25, 4.5) {$u_1$};
		    \node [style=none] (142) at (12.25, 3.5) {$u_2$};
		    \node [style=none] (143) at (12.25, 2.5) {$u_2$};
		    \node [style=none] (144) at (12.25, 1.5) {$u_3$};
		    \node [style=none] (145) at (10.75, 0.5) {$x$};
		    \node [style=none] (146) at (10.75, -0.5) {$y$};
		    \node [style=none] (147) at (12.25, 0.5) {$u_3$};
		    \node [style=none] (148) at (12.25, -0.5) {$v_1$};
		    \node [style=none] (149) at (11.5, -2.5) {$...$};
		    \node [style=none] (150) at (10.75, -1.5) {$y$};
		    \node [style=none] (151) at (12.25, -1.5) {$v_1$};
		    \node [style=none] (152) at (11.5, -4) {$S_2$};
	    \end{pgfonlayer}
	    \begin{pgfonlayer}{edgelayer}
		    \draw (0.center) to (3.center);
		    \draw (3.center) to (2.center);
		    \draw (2.center) to (1.center);
		    \draw (1.center) to (0.center);
		    \draw (15.center) to (18.center);
		    \draw (18.center) to (17.center);
		    \draw (17.center) to (16.center);
		    \draw (16.center) to (15.center);
		    \draw (37.center) to (40.center);
		    \draw (40.center) to (39.center);
		    \draw (39.center) to (38.center);
		    \draw (38.center) to (37.center);
		    \draw (52.center) to (55.center);
		    \draw (55.center) to (54.center);
		    \draw (54.center) to (53.center);
		    \draw (53.center) to (52.center);
		    \draw [->,style=arrow]
		    	(-2,3)
		    	-- 
		    	node[above,font=\Large,text width=1in,align=center] {\added{(2) \\ Obliv. \\ expand}}
		    	node[below,font=\Large,text width=1in,align=center] {\added{Fig.~\ref{fig:oblivious-expand}}}
		    	(0,3);
		    \draw [->,style=arrow]
		    	(15,3)
		    	-- 
		    	node[above,font=\Large,text width=1in,align=center] {\added{(3) \\ Obliv. \\ expand}}
		    	node[below,font=\Large,text width=1in,align=center] {\added{Fig.~\ref{fig:oblivious-expand}}}
		    	(13,3);
		    \draw (109.center) to (112.center);
		    \draw (112.center) to (111.center);
		    \draw (111.center) to (110.center);
		    \draw (110.center) to (109.center);
		    \draw [->,style=arrow]
		    	(10,3)
		    	-- 
		    	node[above,font=\Large,text width=1in,align=center] {\added{(4) \\ Align \\ table}}
		    	node[below,font=\Large,text width=1in,align=center] {\added{Fig.~\ref{fig:table-alignment}}}
		    	(8,3);
		    \draw [->,style=arrow] (3,-2)--(3.8,-4);
		    \draw [->,style=arrow] (5,-2)--(4.2,-4);
		    \node [below,font=\Large,text width=2in,align=center] at (4,-4.25) {\added{(5) Obtain output from a simple row-by-row join}};
		    \node [below,font=\Large,align=center,text width=1.5in] at (-3.5,-1.5) {\added{(1) Compute group dimensions in $T_1$ and $T_2$ (Fig.~\ref{fig:compute-dims}) (omitted from this figure)}};
	    \end{pgfonlayer}
    \end{tikzpicture}
}

\caption{\textbf{Main idea of the algorithm:} The input tables $T_{1}$ and
$T_{2}$ are expanded to produce $S_{1}$ and $S_{2}$, and $S_{2}$
is aligned to $S_{1}$. \added{The output table is then readily obtained by ``zipping'' together  the $d$ values from $S_1$ and $S_2$.}\label{fig:main-idea}}
\end{figure}

If $j_{1},\dots,j_{t}$ are the unique join attribute \added{values}
appearing at least once in each table, then $T_{1}\bowtie T_{2}$
can be written as the union of $t$ partitions:
\[
T_{1}\bowtie T_{2}={\textstyle \bigcup_{i=1}^{t}}\{(d_{1},d_{2})\mid(j_{i},d_{1})\in T_{1},\ (j_{i},d_{2})\in T_{2}\}.
\]
Each partition (henceforth referred to as\emph{ group}) corresponds
to a Cartesian product on sets of size $\alpha_{1}(j_{i})=\mbox{|\{(\ensuremath{j_{i}},\ensuremath{d_{1}})\ensuremath{\in T_{1}}\}|}$
and $\alpha_{2}(j_{i})=\mbox{|\{(\ensuremath{j_{i}},\ensuremath{d_{2}})\ensuremath{\in T_{2}}\}|,}$
respectively, which we call the \emph{dimensions }of the group.

Each entry $(j_{i},d_{1})\in T_{1}$, needs to be matched with $\alpha_{2}(j_{i})$
elements in $T_{2}$; similarly each element $(j_{i},d_{2})\in T_{2}$
needs to be matched with $\alpha_{1}(j_{i})$ elements in $T_{1}$.
To this end, and in similar vein to the work of Arasu et al. \cite{arasu2013oblivious},
we form two \emph{expanded }tables $S_{1}$ and $S_{2}$ (this terminology
is borrowed from their paper), each of size $m=|T_{1}\bowtie T_{2}|$,
such that there are $\alpha_{2}(j_{i})$ copies in $S_{1}$ of each
element $(j_{i},d_{1})\in T_{1}$ and $\alpha_{1}(j_{i})$ copies
in $S_{2}$ of each element $(j_{i},d_{2})\in T_{2}$. Once the expanded
tables are obtained, it only remains to reorder $S_{2}$ to align
with $S_{1}$ so that each copy of $(j_{i},d_{2})\in T_{2}$ appears
at indices in $S_{2}$ that align with each of its $\alpha_{1}(j_{i})$
matching elements from $T_{1}$. At this point, obtaining the final
output is simply a matter of iterating through both tables simultaneously
and collecting the $d$ values from each pair of rows (see \figref{main-idea}).

The approach we use to obtain the expanded tables is very simple,
relying on an oblivious primitive that sends elements to specified
distinct indices in a destination array. Namely, to expand a table
$T$ to $S$ we will first \emph{obliviously distribute} each entry
of $T$ to the index in $S$ where it ought to first occur; this is
achieved by sorting the entries in $T$ by their destination index
and then performing $O(m\log m)$ data-independent swaps so that the
entries ``trickle down'' to their assigned indices. To complete
the expansion, we then perform a single linear pass through the resulting
array to duplicate each non-null entry to the empty slots (containing
null entries) that succeed it.

\section{Algorithm Description}

\begin{algorithm}
\begin{algorithmic}[1]

\Function {Oblivious-Join}{$T_1(j,d), T_2(j,d)$}
	\State $T_1, T_2(j,d,\alpha_1,\alpha_2) \gets \textsc{Augment-Tables}(T_1, T_2)$
	\State $S_1(j, d, \alpha_1, \alpha_2) \gets \textsc{Oblivious-Expand}(T_1, \alpha_2)$
	\State $S_2(j, d, \alpha_1, \alpha_2) \gets \textsc{Oblivious-Expand}(T_2, \alpha_1)$
	\State $S_2 \gets \textsc{Align-Table}(S_2)$
	\State initialize $T_D(d_1, d_2)$ of size $|S_1| = |S_2| = m$
	\For {$i \gets 1 \dots m$}
		\State $T_D[i].d_1 \gets S_1[i].d$
		\State $T_D[i].d_2 \gets S_2[i].d$
	\EndFor
	\State \Return $T_D$

\EndFunction

\end{algorithmic}

\caption{The full oblivious join algorithm\label{alg:full-alg}}
\end{algorithm}

\label{sec:algorithm-description}The complete algorithm is outlined
in \algref{full-alg}, and its subprocedures are described in the
following subsections. We use the notation $T(a_{1},\dots,a_{l})$
when we want to explicitly list the attributes $a_{1},\dots,a_{l}$
of table $T$, for clarity.

We first call $\textsc{Augment-Tables}$ to augment each of the input
tables with attributes $\alpha_{1}$ and $\alpha_{2}$ corresponding
to group dimensions: this process is described in \subsecref{group-dims}.
Then, as detailed in \subsecref{oblivious-expansion}, we obliviously
expand $T_{1}$ and $T_{2}$ into two tables $S_{1}$ and $S_{2}$
of size $m$ each: namely, $S_{1}$ will consist of $\alpha_{2}$
(contiguous) copies of each entry $(j,d_{1},\alpha_{1},\alpha_{2})\in T{}_{1}$,
and likewise $S_{2}$ will consist of $\alpha_{1}$ copies of each
entry $(j,d_{1},\alpha_{1},\alpha_{2})\in T{}_{2}$. To achieve this,
we rely on the oblivious primitive, $\textsc{Oblivious-Distribute}$,
which is the focus of \subsecref{oblivious-distribute}. After expanding
both tables, we call $\textsc{Align-Table}$ to align $S_{2}$ with
$S_{1}$ (with the help of the $\alpha_{1}$ and $\alpha_{2}$ values
stored in $S_{2}$): this amounts to properly ordering $S_{2}$, as
described in \subsecref{table-alignment}. Finally, we collect the
$d$ values from matching rows in $S_{1}$ and $S_{2}$ to obtain
the output table $T_{D}$.

\subsection{Obtaining Group Dimensions\label{subsec:group-dims}}

\begin{algorithm}[!h]
\begin{algorithmic}[1]

\Function {Augment-Tables}{$T_1, T_2$} \Comment{$O(n\log^2 n)$}
	\State $T_C(j,d,tid) \gets (T_1 \times \{tid=1\}) \cup (T_2 \times \{tid=2\})$
	    \State $T_C \gets \textsc{Bitonic-Sort} \langle j\uparrow, tid\uparrow \rangle (T_C)$
		\State $T_C(j, d, tid, \alpha_1, \alpha_2) \gets \textsc{Fill-Dimensions}(T_C)$
	    \State $T_C \gets \textsc{Bitonic-Sort} \langle tid\uparrow, j\uparrow, d\uparrow \rangle (T_C)$
		\State $T_1(j, d, \alpha_1, \alpha_2) \gets T_C[1 \dots n_1]$
		\State $T_2(j, d, \alpha_1, \alpha_2) \gets T_C[n_1+1 \dots n_1+n_2]$
		\State \Return $T_1, T_2$
\EndFunction

\end{algorithmic}

\caption{Augment the tables $T_{1}$ and $T_{2}$ with the dimensions $\alpha_{1}$
and $\alpha_{2}$ of each entry's corresponding group. The resulting
tables are sorted lexicographically by $(j,d)$. $n_{1}=|T_{1}|$,
$n_{2}=|T_{2}|$, $n=n_{1}+n_{2}$. \label{alg:augment-tables}}
\end{algorithm}
\begin{figure}
\begin{centering}
\include{images/3_degs}
\par\end{centering}
\caption{\textbf{Example group dimension calculation.} The dimensions of each
group can be computed by storing temporary counts \deleted{of the
number of entries with the same ID }during a forward pass through
$T_{C}$, and then propagating the totals backwards.\label{fig:compute-dims}}
\end{figure}

Before we expand the two input tables, we need to augment them with
the $\alpha_{1}(j_{i})$ and $\alpha_{2}(j_{i})$ values corresponding
to each join \modified{attribute}{value} $j_{i}$, storing these
in each entry that matches $j_{i}$ (\algref{augment-tables}). To
this end, we need to group all entries with common join \modified{attributes}{values}
together into contiguous blocks, further grouping by them by their
table ID. This is achieved by concatenating both tables (augmented
with table IDs) together and sorting the result lexicographically
by $(j,tid)$, thus obtaining a table $T_{C}$ of size $n=n_{1}+n_{2}$. 

The $\alpha_{1}$ and $\alpha_{2}$ values for each group can then
be obtained by counting the number of entries originating from table
1 and table 2, respectively. Since such entries appear in contiguous
blocks after the sort, this is a matter of keeping count of all entries
with the same ID and storing these counts within all entries of the
same group; in this manner, we can compute all $\alpha_{1}$ and $\alpha_{2}$
values in two linear passes through $T$ (one forward and one backward),
as detailed in $\textsc{Fill-Dimensions}$ and shown in \figref{compute-dims}.
\added{Note that by keeping a sum of the products $\alpha_{1}\alpha_{2}$,
we also obtain the output size $m$, which is needed in subsequent
stages.}

Take for example the join \modified{attribute}{value} $x$, which
corresponds to a group with dimensions $\alpha_{1}=2$ and $\alpha_{2}=3$
(since these are the number of entries with ID 1 and 2, respectively).
While encountering entries with $j=x$ and $tid=1$ during the forward
pass, we temporarily store in the $\alpha_{1}$ attribute of each
entry an incremental count of all previously encountered entries with
$j=x$. When we reach entries with $tid=2$, we can propagate the
final count $\alpha_{1}=2$ to all these entries, while starting a
new incremental count, stored in $\alpha_{2}$. After iterating through
the whole table $T_{C}$ in this manner, $T_{C}$ holds corrects $\alpha_{1}$
and $\alpha_{2}$ values in each ``boundary'' entry (the last entry
within a group, such as $(x,u_{3},\ldots)$), which can then be propagated
to all remaining entries within the same group by iterating through
$T_{C}$ backwards.

It remains for us to extract the augmented $T_{1}$ and $T_{2}$ from
$T_{C}$: to accomplish this, we re-sort $T_{C}$ lexicographically
by $(tid,j,d)$: the first $n_{1}$ values of $T_{C}$ then correspond
to $T_{1}$ (augmented and sorted lexicographically by $(j,d)$),
the remaining $n_{2}$ values correspond to $T_{2}$.

\subsection{Oblivious Distribution\label{subsec:oblivious-distribute}}

\begin{figure}
\begin{centering}
\resizebox{1\columnwidth}{!}{
    \begin{tikzpicture}
        \tikzstyle{every node}=[font=\huge]
	    \begin{pgfonlayer}{nodelayer}
		    \node [style=none] (0) at (-5.5, 6) {};
		    \node [style=none] (1) at (-5.5, 1) {};
		    \node [style=none] (2) at (-2.5, 1) {};
		    \node [style=none] (3) at (-2.5, 6) {};
		    \node [style=none] (4) at (-4.75, 5.5) {$x_1$};
		    \node [style=none] (5) at (-4.75, 4.5) {$x_2$};
		    \node [style=none] (6) at (-4.75, 3.5) {$x_3$};
		    \node [style=none] (7) at (-4.75, 2.5) {$x_4$};
		    \node [style=none] (8) at (-4.75, 1.5) {$x_5$};
		    \node [style=none] (9) at (-3.25, 5.5) {$4$};
		    \node [style=none] (10) at (-3.25, 4.5) {$1$};
		    \node [style=none] (11) at (-3.25, 3.5) {$3$};
		    \node [style=none] (12) at (-3.25, 2.5) {$8$};
		    \node [style=none] (13) at (-3.25, 1.5) {$6$};
		    \node [style=none] (14) at (-4, 0) {$X$};
		    \node [style=none] (15) at (0.5, 6) {};
		    \node [style=none] (16) at (0.5, -2) {};
		    \node [style=none] (17) at (5, -2) {};
		    \node [style=none] (18) at (5, 6) {};
		    \node [style=none] (19) at (1.25, 5.5) {$1$};
		    \node [style=none] (20) at (1.25, 4.5) {$2$};
		    \node [style=none] (21) at (1.25, 3.5) {$3$};
		    \node [style=none] (22) at (1.25, 2.5) {$4$};
		    \node [style=none] (23) at (1.25, 1.5) {$5$};
		    \node [style=none] (24) at (2.75, 5.5) {$x_2$};
		    \node [style=none] (25) at (2.75, 4.5) {$x_3$};
		    \node [style=none] (26) at (2.75, 3.5) {$x_1$};
		    \node [style=none] (27) at (2.75, 2.5) {$x_5$};
		    \node [style=none] (28) at (2.75, 1.5) {$x_4$};
		    \node [style=none] (29) at (2.75, -3) {$A$};
		    \node [style=none] (30) at (-4.75, 6.75) {$x$};
		    \node [style=none] (31) at (-3.25, 6.75) {$f(x)$};
		    \node [style=none] (32) at (1.25, 6.75) {};
		    \node [style=none] (33) at (1.25, 6.75) {$i$};
		    \node [style=none] (34) at (2.75, 6.75) {};
		    \node [style=none] (35) at (2.75, 6.75) {$x$};
		    \node [style=none] (36) at (-1.5, 3.5) {};
		    \node [style=none] (38) at (-0.5, 3.5) {};
		    \node [style=none] (39) at (1.25, 0.5) {$6$};
		    \node [style=none] (40) at (1.25, -0.5) {$7$};
		    \node [style=none] (41) at (1.25, -1.5) {$8$};
		    \node [style=none] (42) at (2.75, 0.5) {-};
		    \node [style=none] (43) at (2.75, -0.5) {-};
		    \node [style=none] (44) at (2.75, -1.5) {-};
		    \node [style=none] (72) at (4.25, 5.5) {$1$};
		    \node [style=none] (73) at (4.25, 4.5) {$3$};
		    \node [style=none] (74) at (4.25, 3.5) {$4$};
		    \node [style=none] (75) at (4.25, 2.5) {$6$};
		    \node [style=none] (76) at (4.25, 1.5) {$8$};
		    \node [style=none] (77) at (4.25, 0.5) {-};
		    \node [style=none] (78) at (4.25, -0.5) {-};
		    \node [style=none] (79) at (4.25, -1.5) {-};
		    \node [style=none] (81) at (4.25, 6.75) {$f(x)$};
		    \node [style=none] (82) at (10.5, 6) {};
		    \node [style=none] (83) at (10.5, -2) {};
		    \node [style=none] (84) at (15, -2) {};
		    \node [style=none] (85) at (15, 6) {};
		    \node [style=none] (86) at (11.25, 5.5) {$1$};
		    \node [style=none] (87) at (11.25, 4.5) {$2$};
		    \node [style=none] (88) at (11.25, 3.5) {$3$};
		    \node [style=none] (89) at (11.25, 2.5) {$4$};
		    \node [style=none] (90) at (11.25, 1.5) {$5$};
		    \node [style=none] (91) at (12.75, 5.5) {$x_2$};
		    \node [style=none] (92) at (12.75, 4.5) {-};
		    \node [style=none] (93) at (12.75, 3.5) {$x_3$};
		    \node [style=none] (94) at (12.75, 2.5) {$x_1$};
		    \node [style=none] (95) at (12.75, 1.5) {-};
		    \node [style=none] (96) at (12.75, -3) {$A$};
		    \node [style=none] (97) at (11.25, 6.75) {};
		    \node [style=none] (98) at (11.25, 6.75) {$i$};
		    \node [style=none] (99) at (12.75, 6.75) {};
		    \node [style=none] (100) at (12.75, 6.75) {$x$};
		    \node [style=none] (101) at (11.25, 0.5) {$6$};
		    \node [style=none] (102) at (11.25, -0.5) {$7$};
		    \node [style=none] (103) at (11.25, -1.5) {$8$};
		    \node [style=none] (104) at (12.75, 0.5) {$x_5$};
		    \node [style=none] (105) at (12.75, -0.5) {-};
		    \node [style=none] (106) at (12.75, -1.5) {$x_4$};
		    \node [style=none] (107) at (14.25, 5.5) {$1$};
		    \node [style=none] (108) at (14.25, 4.5) {-};
		    \node [style=none] (109) at (14.25, 3.5) {$3$};
		    \node [style=none] (110) at (14.25, 2.5) {$4$};
		    \node [style=none] (111) at (14.25, 1.5) {-};
		    \node [style=none] (112) at (14.25, 0.5) {$6$};
		    \node [style=none] (113) at (14.25, -0.5) {-};
		    \node [style=none] (114) at (14.25, -1.5) {$8$};
		    \node [style=none] (116) at (14.25, 6.75) {$f(x)$};
		    \node [style=none] (117) at (8.5, 5.5) {};
		    \node [style=none] (118) at (8.5, 4.5) {};
		    \node [style=none] (119) at (8.5, 3.5) {};
		    \node [style=none] (120) at (8.5, 2.5) {};
		    \node [style=none] (121) at (8.5, 1.5) {};
		    \node [style=none] (122) at (8.5, 0.5) {};
		    \node [style=none] (123) at (8.5, -0.5) {};
		    \node [style=none] (124) at (8.5, -1.5) {};
		    \node [style=none] (125) at (10, 5.5) {};
		    \node [style=none] (126) at (10, 4.5) {};
		    \node [style=none] (127) at (10, 3.5) {};
		    \node [style=none] (128) at (10, 2.5) {};
		    \node [style=none] (129) at (10, 1.5) {};
		    \node [style=none] (130) at (10, 0.5) {};
		    \node [style=none] (131) at (10, -0.5) {};
		    \node [style=none] (132) at (10, -1.5) {};
		    \node [style=none] (133) at (7, 5.5) {};
		    \node [style=none] (134) at (7, 4.5) {};
		    \node [style=none] (135) at (7, 3.5) {};
		    \node [style=none] (136) at (7, 2.5) {};
		    \node [style=none] (137) at (7, 1.5) {};
		    \node [style=none] (138) at (7, 0.5) {};
		    \node [style=none] (139) at (7, -0.5) {};
		    \node [style=none] (140) at (7, -1.5) {};
		    \node [style=none] (141) at (5.5, 5.5) {};
		    \node [style=none] (142) at (5.5, 4.5) {};
		    \node [style=none] (143) at (5.5, 3.5) {};
		    \node [style=none] (144) at (5.5, 2.5) {};
		    \node [style=none] (145) at (5.5, 1.5) {};
		    \node [style=none] (146) at (5.5, 0.5) {};
		    \node [style=none] (147) at (5.5, -0.5) {};
		    \node [style=none] (148) at (5.5, -1.5) {};
	    \end{pgfonlayer}
	    \begin{pgfonlayer}{edgelayer}
		    \draw (0.center) to (3.center);
		    \draw (3.center) to (2.center);
		    \draw (2.center) to (1.center);
		    \draw (1.center) to (0.center);
		    \draw (15.center) to (18.center);
		    \draw (18.center) to (17.center);
		    \draw (17.center) to (16.center);
		    \draw (16.center) to (15.center);
		    \draw [style=arrow] 
		    	(-2.5, 3.5) to
		    	node [above,font=\Large,text width=1in,align=center] {\added{(1) Sort \\ on $f(x)$,}}
		    	node [below,font=\Large,text width=1in,align=center] {\added{add empty rows}}
		    	(0.5, 3.5);
		    \draw (82.center) to (85.center);
		    \draw (85.center) to (84.center);
		    \draw (84.center) to (83.center);
		    \draw (83.center) to (82.center);
		    \draw [style=dotted] (133.center) to (119.center);
		    \draw [style=dotted] (134.center) to (120.center);
		    \draw [style=dotted] (135.center) to (121.center);
		    \draw (136.center) to (122.center);
		    \draw (137.center) to (123.center);
		    \draw [style=dotted] (138.center) to (124.center);
		    \draw (119.center) to (128.center);
		    \draw [style=dotted] (120.center) to (129.center);
		    \draw [style=dotted] (121.center) to (130.center);
		    \draw [style=dotted] (122.center) to (131.center);
		    \draw (123.center) to (132.center);
		    \draw (118.center) to (127.center);
		    \draw [style=dotted] (117.center) to (126.center);
		    \draw (133.center) to (117.center);
		    \draw (117.center) to (125.center);
		    \draw (134.center) to (118.center);
		    \draw (135.center) to (119.center);
		    \draw [style=dotted] (119.center) to (127.center);
		    \draw [style=dotted] (118.center) to (126.center);
		    \draw [style=dotted] (136.center) to (120.center);
		    \draw [style=dotted] (120.center) to (128.center);
		    \draw [style=dotted] (137.center) to (121.center);
		    \draw [style=dotted] (121.center) to (129.center);
		    \draw [style=dotted] (138.center) to (122.center);
		    \draw (122.center) to (130.center);
		    \draw [style=dotted] (139.center) to (123.center);
		    \draw [style=dotted] (123.center) to (131.center);
		    \draw [style=dotted] (140.center) to (124.center);
		    \draw [style=dotted] (124.center) to (132.center);
		    \draw [in=180, out=0] (141.center) to (133.center);
		    \draw [style=dotted] (141.center) to (137.center);
		    \draw [style=dotted] (142.center) to (138.center);
		    \draw [style=dotted] (143.center) to (139.center);
		    \draw [style=dotted] (144.center) to (140.center);
		    \draw (142.center) to (134.center);
		    \draw (143.center) to (135.center);
		    \draw (144.center) to (136.center);
		    \draw (145.center) to (137.center);
		    \draw [style=dotted] (146.center) to (138.center);
		    \draw [style=dotted] (147.center) to (139.center);
		    \draw [style=dotted] (148.center) to (140.center);
		    \node [font=\Large,text width=2in,align=center] at (7.75,-3) {\added{(2) Obliviously \\ route each $x$ \\ to $f(x)$}};
		\end{pgfonlayer}
    \end{tikzpicture}
}
\par\end{centering}
\caption{\textbf{Example oblivious distribution with $n=5$ and $m=8$.} The
intermediate step involves sorting the elements by their target indices.
\modified{Afterwards, through a series of oblivious swaps, the elements
gradually \textquotedblleft gravitate\textquotedblright{} towards their
destinations.}{The elements are then passed through a routing network, which for $m=8$ has hop intervals of size 4, 2, and 1.}
\label{fig:oblivious-distribute}}
\end{figure}
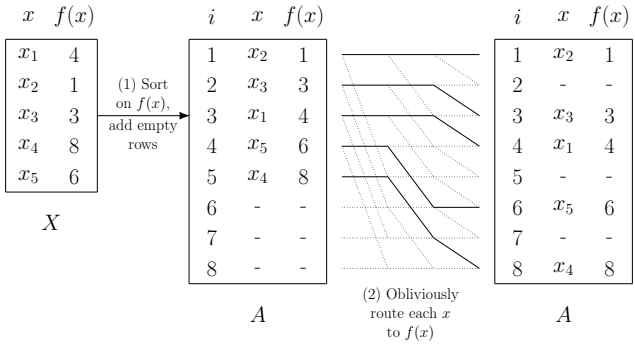

We will reduce expansion to a slightly generalized version of the
following problem: given an input $X=(x_{1},\dots,x_{n})$ of $n$
elements each indexed by an injective map $f:X\to\{1,\dots,m\}$ where
$m\geq n$, the goal of \textsc{Oblivious-Distribute} (as visualized
in \figref{oblivious-distribute}) is to store element $x_{i}$ at
index $f(x_{i})$ of an array $A$ of size $m$. Note that for $m=n$,
the problem is equivalent to that of sorting obliviously; however
for $m>n$, we cannot directly use a sorting network since the output
$A$ needs to contain $m-n$ elements that are not part of the output
(such as placeholder $\varnothing$ values), and we do not know what
indices to assign to such elements so that the $x_{i}$ appear at
their target locations after sorting.

One approach to this problem is probabilistic and requires us to first
compute a pseudorandom permutation $\pi$ of size $m$. We scan through
the $n$ elements, storing element $x_{i}$ at index $\pi(f(x_{i}))$
of $A$. We then use a bitonic sorter to sort the $m$ elements of
$A$ by increasing values of $\pi^{-1}$ applied to each element's
index in $A$. This has the effect of ``undoing'' the masking effect
of the permutation $\pi$ so that if $x_{i}$ is stored at index $\pi(f(x_{i}))$
of $A$, then as soon as $A$ is sorted, it will appear in its correct
destination at index $f(x_{i})$ of $A$. An adversary observing the
accesses of this procedure observes writes at locations $\pi(f(x_{1})),\dots,\pi(f(x_{n}))$
of $A$, followed by the input-independent accesses of the bitonic
sorter. Since $f$ is injective, $f(x_{1}),\dots,f(x_{n})$ are distinct
and so the values $\pi(f(x_{1})),\dots,\pi(f(x_{n}))$ will correspond
to a uniformly-random $n$-sized subset of $\{1,\dots,m\}$. This
approach is therefore oblivious in the probabilistic sense.

\begin{algorithm}
\begin{algorithmic}[1]
\Function {Oblivious-Distribute}{$X, f, m$}
	\State $A[1\ldots n] \gets X$
    \State $\textsc{Bitonic-Sort} \langle f\uparrow \rangle (A)$ \Comment{O($n \log^2 n$)}
	\State $A[n + 1 \dots m] \gets \text{$\varnothing$ values}$
	\State \text{extend $f$ to $\hat{f}$ such that $\hat{f} (\varnothing)=0$}
	\State $j \gets 2^{\lceil \log_2 m \rceil - 1}$
	\While {$j \geq 1$} \Comment{$O(m\log m)$}
		\For {$i \gets m - j \dots 1$}
				\State $y \overset{\star}{\gets} A[i]$
				\State $y' \overset{\star}{\gets} A[i+j]$
				\If {$\hat{f}(y) \geq i + j$}
					\State $A[i] \overset{\star}{\gets} y'$
					\State $A[i+j] \overset{\star}{\gets} y$
				\Else
					\State $A[i] \overset{\star}{\gets} y$
					\State $A[i+j] \overset{\star}{\gets} y'$
				\EndIf
		\EndFor
		\State $j \gets j / 2$
	\EndWhile
	\State \Return $A$

\EndFunction

\end{algorithmic}

\caption{Obliviously map each $x\in X$ to index $f(x)$ of an array of size
$m\protect\geq n$, where $f:X\to\{1\dots m\}$ is injective.\label{alg:oblivious-distribute}}
\end{algorithm}

The second approach, which we use in our implementation and outlined
in \algref{oblivious-distribute}, is deterministic and does not require
the use of a pseudorandom permutation, which can be expensive in practice
and also introduces an extra cryptographic assumption. \added{This
method is similar to the routing network used by Goodrich et al. \cite{goodrich2011data}
for tight order-preserving compaction, except here it used in the
reverse direction (instead of compacting elements together it spreads
them out).} It makes the whole algorithm deterministic (making it
easy to empirically test for obliviousness) and has running time $O(n\log^{2}n+m\log m)$:
the sort takes $O(n\log^{2}n)$ time, the outer loop performs $O(\log m)$
iterations, and the inner loop performs $O(m)$. Intuitively, it is
oblivious since the loops do not depend on the values of $A[i]$,
and though the conditional statement statement depends on $f(A[i])$,
both branches make the same accesses to $A$.

\begin{sloppypar}

The idea is to first sort the $x_{i}$ by increasing destination indices
according to $f$ (by using the notation $\mbox{\textsc{Bitonic-Sort}\ensuremath{\langle f\rangle(A)}}$,
we assume that the $f$ value of each element is stored as an attribute).
Each element can then be sent to its destination index by a series
of $O(\log m)$ hops, where each hop corresponds to an interval $j$
that is a power of two. For decreasing values of $j$, we iterate
through $A$ backwards and perform reads and writes to elements $j$
apart. Most of these will be dummy accesses producing no effect; however,
if we encounter an $x_{i}$ such that $x_{i}$ can hop down a distance
of $j$ and not exceed its target index, we perform an actual swap
with the element stored at that location. This will always be a $\varnothing$
element since the non-null elements ahead of $x_{i}$ make faster
progress, as we will formally show. Therefore each $x_{i}$ will make
progress at the values of $j$ that correspond to its binary expansion,
and it will never be the case that it regresses backwards  by virtue
of being swapped with a non-null element that precedes it in $A$.

\end{sloppypar}

In \figref{oblivious-distribute}, for example, each element must
move a distance that for $m=8$, has a binary expansion involving
the numbers $4$, $2$ and $1$. No $x_{i}$ can make a hop of length
$4$ in this example; however, for the next hop length, $2$, element
$x_{4}$ will advance to index $7$, after which element $x_{5}$
will advance to index $6$ (which at this point corresponds to an
empty cell containing a $\varnothing$ value). Finally, for a hop
length of $1$, element $x_{4}$ will advance to index $8$, element
$x_{1}$ will advance to index $4$ and element $x_{3}$ will advance
to index $3$, in that order; at this point all the elements will
be stored at their desired locations.

We deal with correctness in the following theorem:

\newdef{theorem}{Theorem}
\begin{theorem}

If $m>n$ and $f:\{x_{1},\dots,x_{n}\}\to\{1,\dots,m\}$ is injective,
then \textsc{Oblivious-distribute$(X,f,m)$} returns an $A$ such
that for $1\leq i\leq n$, $A[f(x_{i})]=x_{i}$ (and the remaining
elements of $A$ are $\varnothing$ values).

\begin{proof}

Note that after $A$ is initialized, any write to $A$ is either part
of a swap or leaves $A$ unchanged; thus at the end of the procedure,
$A$ is a permutation of its initial elements and therefore still
contains all the $n$ elements of $X$ and $m-n$ $\varnothing$ values.
After $A$ is sorted, its first $n$ elements, $y_{1},\dots,y_{n}$,
are the elements of $X$ sorted by their values under $f$ (their
destination indices). Note that since $f$ is injective, it follows
that for all $i<j$,
\begin{equation}
j-i\leq f(y_{j})-f(y_{i})
\end{equation}

Let $k=\lceil\log_{2}m\rceil-1$ and let $I_{r}(y_{i})$ be the index
of $y_{i}$ at the end of the $r$-th outer iteration (for $0\leq r\leq k+1$
with $r=0$ corresponding to the state at the start of the loop).
We want to show that for all $i$, $I_{k+1}(y_{i})=f(y_{i})$; this
will follow from the following invariant: at the end of the $r$-th
outer iteration, we have that for all $i$,
\[
0\leq f(y_{i})-I_{r}(y_{i})<2^{k+1-r}.
\]

For $r=0$, the left inequality follows from the fact the $y_{i}$
are sorted by their values under $f$ and $f$ is injective with minimum
value equal 1; the right inequality is simply the bound $f(y_{i})-i\leq f(y_{i})\leq m<2^{k+1}$.
Assuming the invariant holds at iteration $r$, we show that it holds
at iteration $r+1$ as well. 

Consider all (non-dummy) swaps that happen at iteration $r+1$ between
$y=y_{i}$ for some $i$ ($y\neq\varnothing$ since $\hat{f}(\varnothing)=0$)
and some element $y'$ at the index $I_{r}(y_{i})+2^{k-r}$. We show
that it must be the case that $y'=\varnothing$. Suppose instead that
$y'=y_{j}$ for some $j\neq i$. Since neither the index of $y_{i}$
or $y_{j}$ has at that point exceeded $I_{r}(y_{i})+2^{k-r}\leq f(y_{i})<f(y_{j})$,
$y_{i}$ and $y_{j}$ must have both moved forwards the same total
distance throughout the first $r$ iterations and therefore $I_{r}(y_{j})-I_{r}(y_{i})=I_{0}(y_{j})-I_{0}(y_{i})=j-i$.
Since we're assuming that this difference shrinks to zero at the $(r+1)$-th
iteration, it must be the case that 
\[
I_{r+1}(y_{i})=I_{r}(y_{i})+2^{k-r}=I'_{r+1}(y_{j})=I_{r}(y_{j}),
\]
where $I'_{r+1}(y_{j})$ is the index of $y_{j}$ just before it is
swapped with $y_{i}$. Therefore
\[
j-i=I_{r}(y_{j})-I_{r}(y_{i})=2^{k-r}.
\]
This contradicts the fact that
\begin{align*}
j-i & \leq f(y_{j})-f(y_{i}) & \text{\text{(by (1))}}\\
 & \leq f(y_{j})-I_{r}(y_{j}) & \text{(\ensuremath{I_{r}(y_{j})=I_{r+1}(y_{i})\leq f(y_{i})})}\\
 & <2^{k-r} & \text{(since \ensuremath{I'_{r+1}(y_{j})=I_{r}(y_{j})})}
\end{align*}
It follows that no two elements $y_{i}$ and $y_{j}$ are ever swapped,
and therefore:
\[
I_{r+1}(y_{i})=\begin{cases}
I_{r}(y_{i})+2^{k-r}, & I_{r}(y_{i})+2^{k-r}\leq f(y_{i})\\
I_{r}(y_{i}), & \text{otherwise.}
\end{cases}
\]

We can now show that for $i<j$
\[
0\leq f(y_{i})-I_{r+1}(y_{i})<2^{k-r}.
\]
If $I_{r}(y_{i})+2^{k-r}\leq f(y_{i})$, then
\[
f(y_{i})-I_{r+1}(y_{i})=f(y_{i})-I_{r}(y_{i})-2^{k-r}\geq0,
\]
and
\begin{align*}
f(y_{i})-I_{r+1}(y_{i}) & =f(y_{i})-I_{r}(y_{i})-2^{k-r}\\
 & <2^{k-r+1}-2^{k-r}\\
 & =2^{k-r}.
\end{align*}
If $I_{r}(y_{i})+2^{k-r}>f(y_{i})$, then
\[
f(y_{i})-I_{r+1}(y_{i})=f(y_{i})-I_{r}(y_{i})\geq0
\]
\[
f(y_{i})-I_{r+1}(y_{i})=f(y_{i})-I_{r}(y_{i})<2^{k-r},
\]
 which finishes the proof of the invariant.

It then follows from the invariant that $I_{k+1}(y_{i})=f(y_{i})$,
and so when the $k+1$ iterations of the outer loop complete, each
$y_{i}$ will appear in its correct index according to $f$. \end{proof}

\end{theorem}

\subsection{Oblivious Expansion\label{subsec:oblivious-expansion}}

\begin{algorithm}[th]
\begin{algorithmic}[1]

\Function {Oblivious-Expand}{$X, g$}
	\State $\triangleright$ obtain $f$ values and distribute according to $f$
	\State $s \gets 1$
	\For {$i \gets 1 \dots n$} \Comment{$O(n)$}
		\State $x \overset{\star}{\gets} X[i]$
		\If {$g(x) = 0$}
			\State mark $x$ as $\varnothing$ 
		\Else
			\State set $f(x) = s$
		\EndIf
		\State $s \gets s + g(x)$
		\State $X[i] \overset{\star}{\gets} x$
	\EndFor
	\State $A \gets \textsc{Ext-Oblivious-Distribute}(X, f, s-1)$
	\State $\triangleright$ fill in missing entries
	\State $px \gets \varnothing$
	\For {$i \gets 1 \dots s-1$} \Comment{$O(m)$}
		\State $x \overset{\star}{\gets} A[i]$
		\If {$x = \varnothing$}
			\State $x \gets px$
		\Else
			\State $px \gets x$
		\EndIf
		\State $A[i] \overset{\star}{\gets} x$
	\EndFor
	\State \Return $A$
\EndFunction

\State

\Function {Ext-Oblivious-Distribute}{$X, f, m$}
	\State $A[1\ldots n] \gets X$
    \State $\textsc{Bitonic-Sort} \langle \neq \varnothing\uparrow, f\uparrow \rangle (A)$ \Comment{$O(n \log^2 n)$}
	\If {$m \geq n$}
		\State $A[n + 1 \dots m] \gets \text{$\varnothing$ values}$
	\EndIf
	\State \text{extend $f$ to $\hat{f}$ such that $\hat{f} (\varnothing)=0$}
	\State \text{continue as in $O(m \log m)$ loop of \algref{oblivious-distribute}...}
	\State \Return $A[1 \dots m]$
\EndFunction

\end{algorithmic}

\caption{Obliviously duplicate each $x\in X$ $g(x)$ times.\label{alg:oblivious-expand}}
\end{algorithm}
\begin{figure}
\begin{centering}
\resizebox{.89\columnwidth}{!}{
    \begin{tikzpicture}
        \tikzstyle{every node}=[font=\huge]
	    \begin{pgfonlayer}{nodelayer}
		    \node [style=none] (0) at (-6, 6) {};
		    \node [style=none] (1) at (-6, 1) {};
		    \node [style=none] (2) at (-2, 1) {};
		    \node [style=none] (3) at (-2, 6) {};
		    \node [style=none] (4) at (-5.25, 5.5) {$x_1$};
		    \node [style=none] (5) at (-5.25, 4.5) {$x_2$};
		    \node [style=none] (6) at (-5.25, 3.5) {$x_3$};
		    \node [style=none] (7) at (-5.25, 2.5) {$x_4$};
		    \node [style=none] (8) at (-5.25, 1.5) {$x_5$};
		    \node [style=none] (9) at (-4, 5.5) {$2$};
		    \node [style=none] (10) at (-4, 4.5) {$3$};
		    \node [style=none] (11) at (-4, 3.5) {$0$};
		    \node [style=none] (12) at (-4, 2.5) {$2$};
		    \node [style=none] (13) at (-4, 1.5) {$1$};
		    \node [style=none] (14) at (-4, 0) {$X$};
		    \node [style=none] (15) at (2, 6) {};
		    \node [style=none] (16) at (2, -2) {};
		    \node [style=none] (17) at (5, -2) {};
		    \node [style=none] (18) at (5, 6) {};
		    \node [style=none] (19) at (2.75, 5.5) {$1$};
		    \node [style=none] (20) at (2.75, 4.5) {$2$};
		    \node [style=none] (21) at (2.75, 3.5) {$3$};
		    \node [style=none] (22) at (2.75, 2.5) {$4$};
		    \node [style=none] (23) at (2.75, 1.5) {$5$};
		    \node [style=none] (24) at (4.25, 5.5) {$x_1$};
		    \node [style=none] (25) at (4.25, 4.5) {-};
		    \node [style=none] (26) at (4.25, 3.5) {$x_2$};
		    \node [style=none] (27) at (4.25, 2.5) {-};
		    \node [style=none] (28) at (4.25, 1.5) {-};
		    \node [style=none] (29) at (3.5, -3) {$A$};
		    \node [style=none] (30) at (-5.25, 6.75) {$x$};
		    \node [style=none] (31) at (-4, 6.75) {$g(x)$};
		    \node [style=none] (32) at (2.75, 6.75) {};
		    \node [style=none] (33) at (2.75, 6.75) {$i$};
		    \node [style=none] (34) at (4.25, 6.75) {};
		    \node [style=none] (35) at (4.25, 6.75) {$x$};
		    \node [style=none] (36) at (-2, 3.5) {};
		    \node [style=none] (37) at (1, 3.5) {};
		    \node [style=none] (38) at (2, 3.5) {};
		    \node [style=none] (39) at (2.75, 0.5) {$6$};
		    \node [style=none] (40) at (2.75, -0.5) {$7$};
		    \node [style=none] (41) at (2.75, -1.5) {$8$};
		    \node [style=none] (42) at (4.25, 0.5) {$x_4$};
		    \node [style=none] (43) at (4.25, -0.5) {-};
		    \node [style=none] (44) at (4.25, -1.5) {$x_5$};
		    \node [style=none] (70) at (5, 3.5) {};
		    \node [style=none] (71) at (9, 3.5) {};
		    \node [style=none] (72) at (-2.75, 5.5) {$1$};
		    \node [style=none] (73) at (-2.75, 4.5) {$3$};
		    \node [style=none] (74) at (-2.75, 3.5) {-};
		    \node [style=none] (75) at (-2.75, 2.5) {$6$};
		    \node [style=none] (76) at (-2.75, 1.5) {$8$};
		    \node [style=none] (77) at (-2.5, 6.75) {$f(x)$};
		    \node [style=none] (78) at (9, 6) {};
		    \node [style=none] (79) at (9, -2) {};
		    \node [style=none] (80) at (12, -2) {};
		    \node [style=none] (81) at (12, 6) {};
		    \node [style=none] (82) at (9.75, 5.5) {$1$};
		    \node [style=none] (83) at (9.75, 4.5) {$2$};
		    \node [style=none] (84) at (9.75, 3.5) {$3$};
		    \node [style=none] (85) at (9.75, 2.5) {$4$};
		    \node [style=none] (86) at (9.75, 1.5) {$5$};
		    \node [style=none] (87) at (11.25, 5.5) {$x_1$};
		    \node [style=none] (88) at (11.25, 4.5) {$x_1$};
		    \node [style=none] (89) at (11.25, 3.5) {$x_2$};
		    \node [style=none] (90) at (11.25, 2.5) {$x_2$};
		    \node [style=none] (91) at (11.25, 1.5) {$x_2$};
		    \node [style=none] (92) at (10.5, -3) {$A$};
		    \node [style=none] (93) at (9.75, 6.75) {};
		    \node [style=none] (94) at (9.75, 6.75) {$i$};
		    \node [style=none] (95) at (11.25, 6.75) {};
		    \node [style=none] (96) at (11.25, 6.75) {$x$};
		    \node [style=none] (97) at (9.75, 0.5) {$6$};
		    \node [style=none] (98) at (9.75, -0.5) {$7$};
		    \node [style=none] (99) at (9.75, -1.5) {$8$};
		    \node [style=none] (100) at (11.25, 0.5) {$x_4$};
		    \node [style=none] (101) at (11.25, -0.5) {$x_4$};
		    \node [style=none] (102) at (11.25, -1.5) {$x_5$};
	    \end{pgfonlayer}
	    \begin{pgfonlayer}{edgelayer}
		    \draw (0.center) to (3.center);
		    \draw (3.center) to (2.center);
		    \draw (2.center) to (1.center);
		    \draw (1.center) to (0.center);
		    \draw (15.center) to (18.center);
		    \draw (18.center) to (17.center);
		    \draw (17.center) to (16.center);
		    \draw (16.center) to (15.center);
		    \draw [style=arrow] 
		    	(36.center) to 
		    	node[above,font=\Large,text width=1.5in,align=center] {\added{(1) Oblivious distribute}}
		    	node[below,font=\Large,text width=1in,align=center] {\added{Fig.~\ref{fig:oblivious-distribute}}}
		    	(38.center);
		    \draw [style=arrow] (70.center) to
		    	node[above,font=\Large,text width=1.5in,align=center] {\added{(2) Fill down}}
		    	(71.center);
		    \draw (78.center) to (81.center);
		    \draw (81.center) to (80.center);
		    \draw (80.center) to (79.center);
		    \draw (79.center) to (78.center);
	    \end{pgfonlayer}
    \end{tikzpicture}
}
\par\end{centering}
\caption{\textbf{Example oblivious expansion.} This is achieved by obliviously
distributing each element to where it ought to first appear and then
scanning through the resulting array to duplicate each entry in the
null slots that follow. \label{fig:oblivious-expand}}

\end{figure}
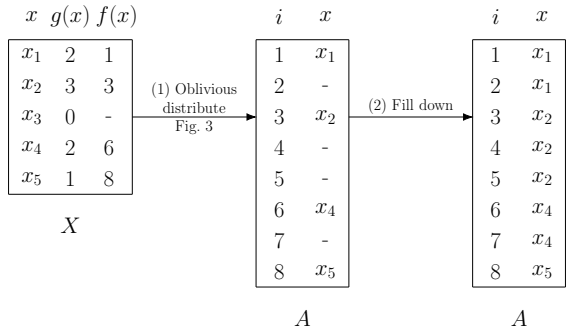

\textsc{Oblivious-Expand} takes an array $X=(x_{1},\dots,x_{n})$
and a function $g$ on $X$ which assigns non-negative integer counts
to each $x$, and outputs
\[
A=(\underbrace{x_{1},\dots,x_{1}}_{\text{\ensuremath{g(x_{1})} times}},\underbrace{x_{2},\dots,x_{2}}_{\text{\ensuremath{g(x_{2})} times}},\dots).
\]
This can easily be achieved using \textsc{Oblivious-Distribute} (see
\figref{oblivious-expand}) if we assume $m\geq n$ and $g(x_{i})>0$
for all $x_{i}$: we compute the cumulative sum $f(x_{i})=1+\sum_{j=1}^{i-1}x_{j},$
and obliviously distribute the $x_{i}$ according to $f$ (in practice,
the values of $f$ are stored as attributes in augmented entries).
The resulting array $A$ is such that each $x_{i}$ is stored in the
first location that it needs to appear in the output of \textsc{Oblivious-Expand};
the next $g(x_{i})-1$ values following $x_{i}$ are all $\varnothing$.
Thus we only need to iterate through $A$, storing the last encountered
entry and using it to overwrite the $\varnothing$ entries that follow.

To account for the possibility that $g(x_{i})=0$ for certain $x_{i}$
(which means that $m$ may possibly be less than $n$), we simply
need to modify \textsc{Oblivious-Distribute }to take as input an $n$-sized
array $X$ such that the subset $X'$ of $X$ of entries not marked
as $\varnothing$ has size $n'\leq m$ and $f':X'\to\{1\ldots m\}$
is injective. The output will be an array $A$ with each $x_{i}\in X'$
stored at index $f(x_{i})$ of $A$; the remainder of $A$ will consist
of $\varnothing$ values as before. This modified version of \textsc{Oblivious-Distribute
(Ext-Oblivious-Distribute) }will allow \textsc{Oblivious-Expand }to
mark entries $x_{i}$ with $g(x_{i})=0$ as $\varnothing$ (done in
practice by first making sure the entries are augmented with an extra
flag bit for this purpose) to the effect that they can be discarded
by \textsc{Ext-Oblivious-Distribute}, as shown in \algref{oblivious-expand}.

\subsection{Table Alignment\label{subsec:table-alignment}}

\begin{algorithm}[t]
\begin{algorithmic}[1]
	\Function {Align-Table}{$S_2$}
		\State $S_2(j,d,\alpha_1,\alpha_2, ii) \gets S_2 \times \{ ii = \texttt{NULL}\}$
		\For {$i \gets 1 \dots |S_2|$}  \Comment{$O(m)$}
			\State $e \overset{\star}{\gets} S_2[i]$
			\State $q \gets \text{(0-based) index of $e$ within block for $e.j$}$
			\State $e.ii \gets \lfloor q / e.\alpha_2 \rfloor + (q \mod  e.\alpha_2)\cdot  e.\alpha_1$
			\State $S_2[i] \overset{\star}{\gets} e$
		\EndFor
		\State $S_2 \gets \textsc{Bitonic-Sort} \langle j, ii \rangle (S_2)$ \Comment{$O(m\log^2 m)$}
		\State \Return $S_2$
	\EndFunction
\end{algorithmic}

\caption{Reorder $S_{2}$ so that its $m$ entries align with those of $S_{1}$.\label{alg:table-alignment}}
\end{algorithm}
\begin{figure}[t]
\begin{centering}
\resizebox{.8\columnwidth}{!}{
    \begin{tikzpicture}
        \tikzstyle{every node}=[font=\huge]
	    \begin{pgfonlayer}{nodelayer}
		    \node [style=none] (37) at (-8, 6) {};
		    \node [style=none] (38) at (-8, -2) {};
		    \node [style=none] (39) at (-5, -2) {};
		    \node [style=none] (40) at (-5, 6) {};
		    \node [style=none] (41) at (-7.25, 5.5) {$x$};
		    \node [style=none] (42) at (-7.25, 4.5) {$x$};
		    \node [style=none] (43) at (-7.25, 3.5) {$x$};
		    \node [style=none] (44) at (-7.25, 2.5) {$x$};
		    \node [style=none] (45) at (-7.25, 1.5) {$x$};
		    \node [style=none] (46) at (-5.75, 5.5) {$a_1$};
		    \node [style=none] (47) at (-5.75, 4.5) {$a_1$};
		    \node [style=none] (48) at (-5.75, 3.5) {$a_1$};
		    \node [style=none] (49) at (-5.75, 2.5) {$a_2$};
		    \node [style=none] (50) at (-5.75, 1.5) {$a_2$};
		    \node [style=none] (51) at (-6.5, -3) {$S_1$};
		    \node [style=none] (52) at (-2, 6) {};
		    \node [style=none] (53) at (-2, -2) {};
		    \node [style=none] (54) at (1, -2) {};
		    \node [style=none] (55) at (1, 6) {};
		    \node [style=none] (66) at (-0.5, -3) {$S_2$};
		    \node [style=none] (73) at (-7.25, 0.5) {$x$};
		    \node [style=none] (74) at (-7.25, -0.5) {$y$};
		    \node [style=none] (76) at (-5.75, 0.5) {$a_2$};
		    \node [style=none] (77) at (-5.75, -0.5) {$b_1$};
		    \node [style=none] (80) at (-6.5, -1.5) {$...$};
		    \node [style=none] (81) at (-1.25, 5.5) {$x$};
		    \node [style=none] (82) at (-1.25, 4.5) {$x$};
		    \node [style=none] (83) at (-1.25, 3.5) {$x$};
		    \node [style=none] (84) at (-1.25, 2.5) {$x$};
		    \node [style=none] (85) at (-1.25, 1.5) {$x$};
		    \node [style=none] (86) at (0.25, 5.5) {$u_1$};
		    \node [style=none] (87) at (0.25, 4.5) {$u_1$};
		    \node [style=none] (88) at (0.25, 3.5) {$u_2$};
		    \node [style=none] (89) at (0.25, 2.5) {$u_2$};
		    \node [style=none] (90) at (0.25, 1.5) {$u_3$};
		    \node [style=none] (91) at (-1.25, 0.5) {$x$};
		    \node [style=none] (92) at (-1.25, -0.5) {$y$};
		    \node [style=none] (94) at (0.25, 0.5) {$u_3$};
		    \node [style=none] (95) at (0.25, -0.5) {$v_1$};
		    \node [style=none] (98) at (-0.5, -1.5) {$...$};
		    \node [style=none] (105) at (-7.25, 6.75) {$j$};
		    \node [style=none] (106) at (-5.75, 6.75) {$d$};
		    \node [style=none] (107) at (-1.25, 6.75) {$j$};
		    \node [style=none] (108) at (0.25, 6.75) {$d$};
		    \node [style=none] (109) at (5, 6) {};
		    \node [style=none] (110) at (5, -2) {};
		    \node [style=none] (111) at (8, -2) {};
		    \node [style=none] (112) at (8, 6) {};
		    \node [style=none] (113) at (6.5, -3) {$S_2$};
		    \node [style=none] (114) at (5.75, 5.5) {$x$};
		    \node [style=none] (115) at (5.75, 4.5) {$x$};
		    \node [style=none] (116) at (5.75, 3.5) {$x$};
		    \node [style=none] (117) at (5.75, 2.5) {$x$};
		    \node [style=none] (118) at (5.75, 1.5) {$x$};
		    \node [style=none] (119) at (7.25, 5.5) {$u_1$};
		    \node [style=none] (120) at (7.25, 4.5) {$u_2$};
		    \node [style=none] (121) at (7.25, 3.5) {$u_3$};
		    \node [style=none] (122) at (7.25, 2.5) {$u_1$};
		    \node [style=none] (123) at (7.25, 1.5) {$u_2$};
		    \node [style=none] (124) at (5.75, 0.5) {$x$};
		    \node [style=none] (125) at (5.75, -0.5) {$y$};
		    \node [style=none] (127) at (7.25, 0.5) {$u_3$};
		    \node [style=none] (128) at (7.25, -0.5) {$v_1$};
		    \node [style=none] (130) at (6.5, -1.5) {$...$};
		    \node [style=none] (131) at (5.75, 6.75) {$j$};
		    \node [style=none] (132) at (7.25, 6.75) {$d$};
		    \node [style=none] (133) at (1, 2) {};
		    \node [style=none] (134) at (5, 2) {};
	    \end{pgfonlayer}
	    \begin{pgfonlayer}{edgelayer}
		    \draw (37.center) to (40.center);
		    \draw (40.center) to (39.center);
		    \draw (39.center) to (38.center);
		    \draw (38.center) to (37.center);
		    \draw (52.center) to (55.center);
		    \draw (55.center) to (54.center);
		    \draw (54.center) to (53.center);
		    \draw (53.center) to (52.center);
		    \draw (109.center) to (112.center);
		    \draw (112.center) to (111.center);
		    \draw (111.center) to (110.center);
		    \draw (110.center) to (109.center);
		    \draw [style=arrow] (133.center) to 
		    	node[above,font=\Large,text width=1.25in,align=center] {\added{Reorder $d$ values within each group of $S_2$.}}
		    	(134.center);
	    \end{pgfonlayer}
    \end{tikzpicture}
}
\par\end{centering}
\caption{\textbf{Example table alignment.} $S_{2}$ is reordered to align with
$S_{1}$. \added{In this example, each of the two copies of $(x,u_{1})$
in $S_{2}$ ends up appearing at two indices matching both $(x,a_{1})$
and $(x,a_{2})$ from $S_{1}$; the same applies to the copies of
$(x,u_{2})$ and $(x,u_{3})$. } \label{fig:table-alignment}}

\end{figure}
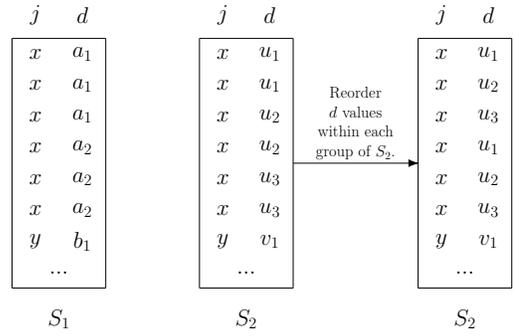

Recall that $S_{1}$ is obtained from $T_{1}$ based on the counts
stored in $\alpha_{2}$ since for each entry $(j_{i},d_{1})\in T_{1}$,
$\alpha_{2}(j_{i})$ is the number of entries in $T_{2}$ \modified{with
attribute}{matching} $j_{i}$, and these are all the entries that
$(j_{i},d_{1})$ must be matched with. Likewise $S_{2}$ is obtained
from $T_{2}$ based on the counts stored in $\alpha_{1}$. It remains
for us to properly align $S_{2}$ to $S_{1}$ so that each output
entry corresponds to a row of $S_{1}$ and a row of $S_{2}$ with
matching index. More precisely, we need to sort $S_{2}$ so that the
sequence of pairs $\{(S_{1}[i].d_{1},S_{2}[i].d_{2})\}_{i=1}^{m}$
is a lexicographic ordering of all the pairs in $T_{1}\bowtie T_{2}$.
For example, in \figref{table-alignment}, the $\alpha_{2}(x)=2$
copies of $(x,u_{1})$ in $S_{2}$, need to be matched with $\alpha_{2}(x)=2$
entries from $T_{1}$: $(x,a_{1})$ and $(x,a_{2})$. Since the entries
in $S_{1}$ occur in blocks of size $\alpha_{1}(x)=3$, this means
that the copies of $(x,u_{1})$ in $S_{2}$ need to occur a distance
of $\alpha_{1}(x)=3$ apart: at indices $1$ and $4$ in $S_{2}$.
In general, these indices can be computed from the $\alpha_{1}$ and
$\alpha_{2}$ attributes, as outlined in \algref{table-alignment}.
Note that $q$ is simply a counter that is reset when a new join \modified{attribute}{value}
is encountered, similarly to the counter $c$ in \algref{augment-tables}.\pagebreak{}

\section{Evaluation}

\label{sec:evaluation}

\modified{We implemented a (sequential) C++ prototype (available
at https://git.uwaterloo.ca/skrastni/obliv-join-impl) of the general
algorithm that is not bound to any particular hardware setup. The
prototype takes as input a plaintext file consisting of two tables
specified by $(j,d)$ pairs (of integers), computes the join in the
manner  described, and outputs the resulting plaintext consisting
of $(d_{1},d_{2})$ pairs. All contents of (heap-allocated) memory
that correspond to public memory --- all except a constant number
of variables such as counters and those used to store the results
of a constant number of read entries --- are accessed through a wrapper
class which is used to keep a log of such accesses. 

We empirically tested the correctness of the implementation by generating
a large number of tests ($\approx$ 1,000) corresponding to different
values of $n$ (from 10 to 1,000,000), and the wide variety of join
groups an input of size $n$ could possibly induce (for instance,
$n$ $1\times1$ groups, a single $1\times n$ group, or groups of
different sizes whose widths and heights are drawn from distributions
with different parameters). Our implementation produced correct outputs
in all the cases.}{

\begin{sloppypar} We implemented a (sequential) C++ prototype of
the general algorithm, which we then readily adapted as an SGX application
whose entire execution takes place within the enclave (code available
at \url{https://git.uwaterloo.ca/skrastni/obliv-join-impl}). We empirically
tested for correctness on varying input sizes $n$ (10 to 1,000,000):
for each $n$, we automatically generated 20 tests consisting of various
different inputs of size $n$ (for instance, one inducing $n$ $1\times1$
groups, one inducing a single $1\times n$ group, and several where
the group sizes were drawn from a power law distribution). The outputs
were correct in all the cases. \end{sloppypar} }

\subsection{Security Analysis\label{subsec:security-analysis}}

\modified{We verified the obliviousness of our implementation both
formally, through the use of a dedicated type system, and empirically,
by comparing the logs of memory accesses for different inputs. }{We
verified the obliviousness of our prototype both formally, through
the use of a dedicated type system, and empirically, by comparing
the logs of array accesses for different inputs. To ensure that that
the actual low-level memory accesses were also oblivious, we transformed
it as per \subsecref{obliv-to-circuit} and inspected its accesses
using an instrumentation tool. }

\subsubsection*{Verification of Obliviousness through Typing}

\begin{figure}[!t]
\begin{gather*}
\inference[T-Var]{\Gamma(x)=\text{Var}\ l}{\Gamma\vdash x:\text{Var}\ l;\ \epsilon}\quad\inference[T-Const]{}{\Gamma\vdash\text{Var L};\ \epsilon}\\
\\
\inference[T-Op]{\Gamma\vdash x:\text{Var}\ l_{1};\ \epsilon\quad\Gamma\vdash y:\text{Var}\ l_{2}:\ \epsilon}{\Gamma\vdash x\ op\ y:\text{Var}\ l_{1}\sqcup l_{2};\ \epsilon}\\
\\
\inference[T-Asgn]{\Gamma(x)=\text{Var}\ l_{1};\ \epsilon\quad\Gamma\vdash y:\text{Var}\ l_{2};\ \epsilon\quad l_{2}\sqsubseteq l_{1}}{\Gamma\vdash x\gets y;\ \epsilon}\\
\\
\Gamma(y)=\text{Arr}\ l'\quad l'\sqsubseteq l\\
\inference[T-Read]{\Gamma\vdash i:\text{Var \ }L;\ \epsilon\quad\Gamma\vdash x:\text{Var}\ l;\ \epsilon}{\Gamma\vdash x\overset{\star}{\gets}y[i];\ \langle R,y,i\rangle}\\
\\
\Gamma(y)=\text{Arr}\ l'\quad l\sqsubseteq l'\\
\inference[T-Write]{\Gamma\vdash i:\text{Var \ }L;\ \epsilon\quad\Gamma\vdash x:\text{Var}\ l;\ \epsilon}{\Gamma\vdash y[i]\overset{\star}{\gets}x;\ \langle W,y,i\rangle}\\
\\
\inference[T-Cond]{\Gamma\vdash c:\text{Var}\ l;\ \epsilon\quad\Gamma\vdash s_{1};\ T\quad\Gamma\vdash s_{2};\ T}{\Gamma\vdash\textbf{\text{if }}c\text{ {\bf then }}s_{1}\text{ {\bf else }}s_{2};\ T}\\
\\
\inference[T-For]{\Gamma\vdash t:\text{Var}\ L;\ \epsilon\quad\Gamma\vdash s;\ T}{\Gamma\vdash\textbf{\text{for }}i\gets1\ldots t\text{ {\bf do }}s;\ \underbrace{T||\ldots||T}_{t\text{ copies}}}\\
\\
\inference[T-Seq]{\Gamma\vdash s_{1};\ T_{1}\quad\Gamma\vdash s_{2};\ T_{2}}{\Gamma\vdash s_{1};s_{2};\ T_{1}||T_{2}}
\end{gather*}

\caption{\textbf{Summary of type system used to model level II obliviousness
and verify implementation.}\label{fig:type-system}}
\end{figure}

Liu et al. \cite{liu2013memory} showed that programming language
techniques can be used to verify the obliviousness of programs. The
authors formally define the concept of memory trace obliviousness
(roughly corresponding to our notion of level III obliviousness),
and define a type system in which only programs satisfying this property
are well-typed. We adapted a simplified version of their system that
does not incorporate the use of ORAM (since we do not use any), and
which corresponds to level II obliviousness in accordance with our
high-level description in the previous section. 

The type system is presented in \figref{type-system}, in a condensed
format. Each type is a pair of the form $\tau;\ T$, where $\tau$
is either $\text{Var}\ l$, $\text{Array}\ l$, or a statement, and
$T$ is a corresponding \emph{trace}. In the case when $\tau$ is
$\text{Var}\ l$ or $\text{Array}\ l$, the label $l$ is either $L$
(``low'' security) if the variable or array stores input-independent
data, or $H$ (``high'' security) otherwise. The ordering relation
on labels, $l_{1}\sqsubseteq l_{2}$, is satisfied when $l_{1}=l_{2}=L$,
or $l_{1}=L$ and $l_{1}=H$. We define $l_{1}\sqcup l_{2}$ to be
$H$ if at least one of $l_{1}$ or $l_{2}$ is $H$ and $L$ otherwise.
In an actual program, we would set to $L$ the label of variables
corresponding to the values of $n$ and $m$, and set to $H$ the
label of all allocated arrays that will contain input-dependent data
(in our program all arrays are such). The trace $T$ is a sequence
of memory accesses $\langle R,y,i\rangle$ (reads) or $\langle W,y,i\rangle$
(writes), where $y$ is the accessed array and $i$ is the accessed
index. We use $\epsilon$ to denote an empty trace and $||$ to denote
the concatenation operator. 

All judgements for expressions are of the form $\Gamma\vdash exp:\tau;\ T,$
where $\Gamma$ is an environment mapping variables and arrays to
types, $exp$ is an expression and $\tau$ is its type, and $T$ is
the trace produced when evaluating $exp$. Judgments for statements
are of the form $\Gamma\vdash s;\ T.$

Note that all rules that involve reads and writes to only Var types
emit no trace since they model our notion of local memory. The rule
T-Asgn models the flow of high-security data: a variable $x$ that
is the target of an assignment involving an $H$ variable $y$ must
always be labeled $H$. The rules T-Read and T-Write are similar to
T-Asgn but also ensure two other properties: that arrays are always
indexed by variables labeled $L$ (for otherwise the memory access
would leak data-dependent data), and that the reads and writes to
arrays emit a trace consisting of the corresponding memory access.
The two rules that play an important role in modeling obliviousness
are T-Cond, which ensures that the two branches of any conditional
statement emit the same memory traces, and T-For, which ensures the
number of iterations of any loop is a low-security variable (such
as a constant, $n$, or $m$). 

We manually verified that our implementation is well-typed in this
system by annotating the code with the correctly inferred types. For
example, every if statement was annotated with the matching trace
of its branches.

\subsubsection*{Experiments: Memory Access Logs}

\begin{figure*}
\includegraphics[width=0.8\paperwidth]{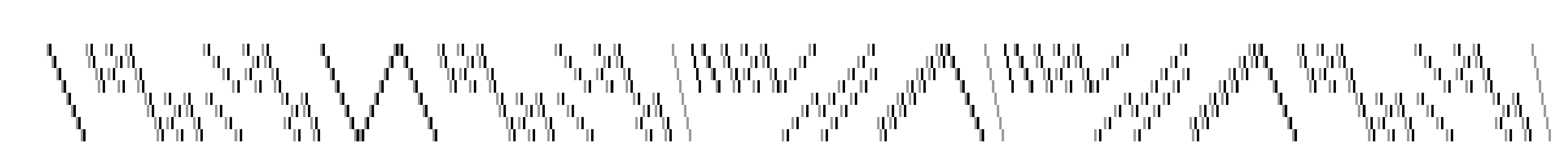}

\caption{\textbf{Visualization of our implementation's input-independent pattern
of memory access as it joins two tables of size 4 into a table of
size 8.} Horizontal axis is (discretized) time, vertical axis is the
memory index; light shade denotes a read; dark denotes a write. \label{fig:memory-accesses}}
\end{figure*}

\added{In our prototype all contents of (heap-allocated) memory that
correspond to public memory --- all except a constant number of variables
such as counters and those used to store the results of a constant
number of read entries --- are accessed through a wrapper class which
is used to keep a log of such accesses.} For small $n$ ($n\leq10)$,
we manually created different test classes (around 5), where each
test class corresponds to values of $n_{1}$ and $n_{2}$ (summing
to $n$), and an output length $m$. We verified, by direct comparison,
that the memory access logs for each of the inputs in the same class
were identical. \figref{memory-accesses} visualizes the full sequence
of memory accesses for $n_{1}=n_{2}=4$ and $m=8$.

For larger values of $n$ where the logs were too large to fit in
memory, we kept a hash of the log instead. That is, we set $H=0$,
and for every access to an index $i$ of an array $A_{r}$ allocated
by our program, we updated $H$ as follows:
\[
H\gets h(H||r||t||i),
\]
where $h$ is a cryptographic hash function (SHA-256 in our case)
and $t$ is $0$ or $1$ depending on whether the access is a read
or a write to $A_{r}$. \modified{With $n$ ranging from 20 to 20,000,
we programatically generated 14 different test classes, with $50$
inputs in each class (generated to represent a diverse range of group
dimensions), and verified that the resulting hashes were the same
for all inputs in the same class. }{With $n$ ranging from 10 to
10,000, we generated a diverse range of tests, in the manner described
at the start of this section, but also under the restriction that
the tests for each $n$ produce outputs of the same size. We verified
that for each $n$ the tests produced the same hash. }

\added{

\subsubsection*{Experiments: Memory Trace Instrumentation}

Through a mix of manual and automated code transformations similar
to those outlined in \subsecref{obliv-to-circuit}, we obtained a
program where all virtual memory accesses of the program are oblivious.
To verify this we ran the same hash-based tests as previously described
except that the target memory accesses were obtained by using Intel's
Pin instrumentation framework to inject the hash computation at every
program instruction involving a memory operand. The verification was
successful when the program was compiled with GCC 7.5.0 with an \texttt{-O2}
optimization level (whereas \texttt{-O3} did not preserve the intended
properties of our transformation).

}

\subsection{Performance Analysis}

\begin{table}
\caption{\textbf{For each (non-linear) component of the algorithm: approximate
counts of total comparisons (or swaps) when $m\approx n_{1}=n_{2}$,
as well as empirical share of total implementation runtime for $n=10^{6}$.}
\label{tab:performance-breakdown}}

\centering{}%
\begin{tabular}{ccc}
\toprule 
\textbf{Subroutine} & \textbf{Comparisons} & \textbf{Runtime}\tabularnewline
\midrule
\multirow{1}{*}{initial sorts on $T_{C}$} & \multirow{1}{*}{$n(\log_{2}n)^{2}/2$} & \multirow{1}{*}{60\%}\tabularnewline
\multirow{1}{*}{o.d. on $T_{1},T_{2}$ (sort)} & $n_{1}(\log_{2}n_{1})^{2}/2$ & \multirow{1}{*}{25\%}\tabularnewline
\multirow{1}{*}{o.d. on $T_{1},T_{2}$ (route)} & \multirow{1}{*}{$2m\log_{2}m$} & \multirow{1}{*}{3\%}\tabularnewline
\multirow{1}{*}{align sort on $S_{2}$} & \multirow{1}{*}{$m(\log_{2}m)^{2}/4$} & \multirow{1}{*}{12\%}\tabularnewline
\midrule 
total & \multirow{2}{*}{$n(\log_{2}n)^{2}+n\log_{2}n$} & \multirow{2}{*}{100\%}\tabularnewline
(when $m\approx n_{1}=n_{2})$ &  & \tabularnewline
\bottomrule
\end{tabular}
\end{table}

Taking into account the fact that performing a bitonic sort on input
$n$ makes roughly $n(\log_{2}n)^{2}/4$ comparisons, the cost breakdown
of the full algorithm is summarized in \tabref{performance-breakdown},
which supports the fact that our time complexity of $O(n\log^{2}n+m\log m)$
does not hide large constants that would make the algorithm impractical.

In terms of space usage, the total (non-oblivious) memory we use is
$\max(n_{1},m)+\max(n_{2},m)$ entries since the table $T_{C}$ has
size $n_{1}+n_{2}$, the augmented tables $T_{1}$ and $T_{2}$ correspond
to two regions of $T_{C}$, and the expanded tables $S_{1}$ and $S_{2}$
can be obtained from $T_{1}$ and $T_{2}$ by only allocating as many
extra entries as needed to expand $T_{1}$ and $T_{2}$ to tables
of size $m$ (if one of the original tables has size less than $m$,
then no extra entries will be allocated for that table's expansion).

\begin{figure}
\begin{tikzpicture}
\begin{axis}[
	height=6cm,
	width=0.47\textwidth,
	xlabel={Input size ($n$)},
	ylabel={Runtime (s)},
	xmin=0,
	ymin=0,
	xmax=1150000,
	xtick = {100000, 250000,500000,750000,1000000},
	xtick style = {draw=none},
	minor y tick num=4,
	tick pos = left,
	legend style={
		at={(0.05,0.85)},
		anchor=west,
		legend cell align=left,
	}
]
	\addplot[color=orange, mark=|] table [x=n,y=sgxl3] {perf_results.csv}
	node[pos=1, label=right:6.30]{};
	\addlegendentry{SGX (transformed)};
	\addplot[color=Green, mark=x] table [x=n,y=sgx] {perf_results.csv}
	node[pos=1, label=right:5.67]{};
	\addlegendentry{SGX};
	\addplot[color=red, mark=diamond] table [x=n,y=prototype] {perf_results.csv}
	node[pos=1, label=right:2.35]{};
	\addlegendentry{prototype};
	\addplot[color=blue, mark=star] table [x=n,y=mergejoin] {perf_results.csv}
	node[pos=1, label=above right:0.03]{};
	\addlegendentry{insecure sort-merge};
\end{axis}
\end{tikzpicture}

\caption{\textbf{Performance results for sequential prototype implementation.
}The inputs are such that $m\approx n_{1}=n_{2}=n/2$.\textbf{ }\label{fig:performance-results}}
\end{figure}
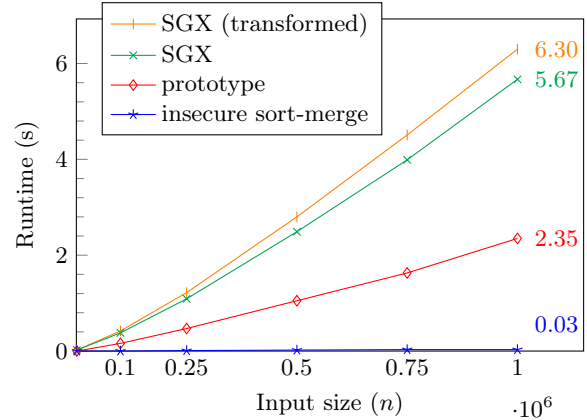

\modified{We ran our prototype implementation on a single core of
an Intel Core i5-7300U 2.60 GHz laptop with 8 GB RAM; the runtime
(both total and that of \textsc{Oblivious-Distribute}) for a selection
of input sizes ranging from $n=1,000$ to $n=1,000,000$ and output
sizes roughly equal to $n/2$ is shown in Figure 8.}{ We ran the
different variants of our implementation on a single core of an Intel
Core i5-7300U 2.60 GHz laptop with 8 GB RAM; the runtime of the prototype,
the SGX version, and the transformed SGX version is shown in \figref{performance-results}
and compared to a non-oblivious sort-merge join. Since our SGX versions
exclusively use the limited Enclave Page Cache (EPC) of size approximately
93 MiB for all allocated memory, we anticipate a drop in performance
for input sizes where the EPC size is insufficient (due to swapping).
However, this size is expected to be increased considerably in future
versions of SGX.

The only related join algorithm with an implementation that has been
evaluated on input sizes up to $10^{6}$ is the one proposed by Opaque,
which we remind is restricted to primary-foreign key joins. Its SGX
implementation, despite being evaluated on better hardware and on
multiple cores, runs approximately five times slower for an input
size of $n=10^{6}$.}

\begin{sloppypar} Although our implementation is non-parallel, almost
all parts of our algorithm are amenable to parallelization since they
heavily rely on sorting networks, whose depth is $O(\log^{2}n)$.
The only exception is the sequence of $O(m\log m)$ operations following
the sorts in each of the two calls to \textsc{Oblivious-Distribute.}
\modified{We have not investigated whether these operations can be
parallelized in a meaningful way.}{However, as is shown in \tabref{performance-breakdown},
these operations account for a negligibly small fraction of the total
runtime.} \end{sloppypar}

\section{Conclusions and Future Work}

Our algorithm for oblivious joins has a runtime that closely approaches
that of the standard sort-merge join and has a low total operation
count. Being based on sorting networks and similar constructions,
it has very low circuit complexity and introduces novel data-independent
techniques for query processing. There is an increasing demand for
such approaches due to their resistance against side-channel attacks
and suitability for secure computation.

We have not yet considered whether compound queries involving joins
\added{(including multi-way joins)} can be \modified{optimized}{readily obtained}
using the techniques for this paper. Grouping aggregations over joins
could be computed using fewer sorting steps than a full join would
require, for example, by combining the work of Arasu et al. \cite{arasu2013oblivious}
in this direction with the primitives we provide. These primitives,
especially oblivious distribution and expansion, could also potentially
be useful in providing a general framework for oblivious algorithm
design or have direct applications in various different problem areas
with similar security goals.

\section{Acknowledgments}

We gratefully acknowledge the support of NSERC for grants RGPIN-2016-05146,
RGPIN-05849, RGPAS-507908, CRDPJ-531191, and DGDND-00085, and the
Royal Bank of Canada for funding this research. 

\balance\newpage\bibliographystyle{abbrv}
\bibliography{obliv_join}

\end{document}